# Traversing double-well potential energy surfaces: photoinduced concurrent intralayer and interlayer structural transitions in $XTe_2$ (X=Mo, W)

Yingpeng Qi[1,3#*], Mengxue Guan[2#], Daniela Zahn[1], Thomas Vasileiadis[1], Hélène Seiler[1], Yoav William Windsor[1], Hui Zhao[2], Sheng Meng[2,4*], Ralph Ernstorfer[1*]

[1]Fritz-Haber-Institut der Max-Planck-Gesellschaft, Faradayweg 4-6, Berlin 14195, Germany.

[2]Beijing National Laboratory for Condensed Matter Physics and Institute of Physics, Chinese Academy of Sciences, Beijing 100190, P. R. China.

[3]Center for Ultrafast Science and Technology, School of Physics and Astronomy, Shanghai Jiao Tong University, 200240 Shanghai, China.

[4]School of Physical Sciences, University of Chinese Academy of Sciences, Beijing 100049, China.

[#]These authors contributed equally: Yingpeng Qi, Meng-Xue Guan

*Correspondence to qiyp@sjtu.edu.cn, smeng@iphy.ac.cn and ernstorfer@fhi-berlin.mpg.de

Manipulating crystal structure and the corresponding electronic properties in quantum materials provides opportunities for the exploration of exotic physics and practical applications. Here, by ultrafast electron diffraction, structure factor calculation and TDDFT-MD simulations, we report the photoinduced concurrent intralayer and interlayer structural transitions in the Td and 1T' phase of $XTe_2$ (X=Mo, W). Concomitant with the interlayer structural transition by shear displacement, the ultrafast suppression of the intralayer Peierls distortion within 0.3 ps is demonstrated and attributed to Mo-Mo (W-W) bond stretching. We discuss the modification of multiple quantum electronic states associated with the intralayer and interlayer structural transitions, such as the topological band inversion and the higher-order topological state. The twin structure and the stacking fault in $XTe_2$ are identified by the ultrafast structural response. Our work elucidates the pathway of the photoinduced intralayer and interlayer structural transitions in atomic and femtosecond spatiotemporal scale. Moreover, the concurrent intralayer and interlayer structural transitions reveals the traversal of all double-well potential energy surfaces (DWPES) by laser excitation in material system, which may be an intrinsic mechanism in the field of photoexcitation-driven symmetry engineering, beyond the single DWPES transition model and the order-disorder transition model.

## Introduction

Within the transition metal dichalcogenides (TMDCs) family, $MoTe_2$ and $WTe_2$ (i.e. $XTe_2$ (X=Mo, W)), have recently sparked broad research interest for their rich structural phases and unusual electronic structures, such as the semiconductor-to-semimetal structural transition (*1*, *2*), the extremely large magnetoresistance (*2, 3*), the quantum spin Hall effect (*4-7*) and the novel topological phases (*8-14*). At room temperature, $MoTe_2$ is in its 1T' phase and forms a layered structure with double sheets of Te atoms bound together by interleaving Mo atoms, as shown in Fig. 1A. By lowering the temperature to below 250 K, bulk $MoTe_2$ undergoes a structural transition from the monoclinic 1T' to the orthorhombic Td phase. These two semimetallic phases hold distinct interlayer stacking while exhibit the same intralayer crystal structure with a Peierls distortion (*2*). The intralayer Peierls distortion is characterized by the Mo-Mo metallic bonds. As shown in Fig. 1A, such exotic Mo-Mo metallic bonds modulate the adjacent Mo atoms by alternating shorter and longer distances and also drive the out-of-plane wrinkling of the Te atoms. The 1T' phase of $MoTe_2$ belongs to the centrosymmetric space group ($P2_1/m$), and the Td phase of $MoTe_2$ belongs to the non-centrosymmetric space group ($Pmn2_1$). As a sister compound of $MoTe_2$, $WTe_2$ has commonly been observed in the Td phase, even in a high temperature. Both the 1T' and Td phase of $XTe_2$ are topological nontrivial. The bulk Td phase of $XTe_2$ is a type-II Weyl semimetal (*8-12*, *15*), meanwhile, the bulk Td phase (and the 1T' phase) is also a higher-order topological insulator (*13*, *14*, *16*). The monolayer of $XTe_2$ is a topological insulator (*2*, *4*, *5*, *17*). The crystalline topological phase of electrons is intrinsically protected by the symmetry of the crystal (*18-21*). Therefore, triggering an intralayer or interlayer symmetry change offers prospects in practical applications such as topological switch electronics.

Regarding the interlayer symmetry change, an interlayer stacking transition with THz field pump is observed for the Td phase of $WTe_2$ (*22*). Consequently, the Weyl semimetal phase is switched to the trivial phase by the coherent interlayer shear phonon mode. With infrared (800 nm) and midinfrared (2600 nm) laser pump (*23*), a similar topological switch in the Td phase of $MoTe_2$ is observed through ultrafast spectroscopy. The recent angle-resolved photoemission spectroscopy (ARPES) study shows that the interlay shear mode can be described by a $\pi$-shifted sine function, implying an impulsive excitation of this shear mode and therefore a model of field driven stacking transition (*24*). However, such a field driven stacking transition model is controversial because the electron-phonon interaction induced by the above-band-gap photon excitation is neglected. A direct observation of the ultrafast interlayer stacking transition with a structural probe is required to identify the mechanism.

In $XTe_2$, the intralayer Peierls distortion brings about the onset of unique physical phenomena, such as unprecedented anisotropic optical and electronic properties (*25*). More importantly, the intralayer distortion induces the band inversion. Such a band inversion causes the monolayer $XTe_2$ to become a topological insulator (*2*, *4*, *5*, *17*) and the bulk to become a higher-order topological insulator (*13*, *14*, *16*). A schematic illustration of the band inversion in monolayer 1T' $MoTe_2$ is shown in Fig. 1B. With the inclusion of spin–orbit coupling (SOC), the band hybridization and the lifting of degeneracies at the Dirac cones (driven by the band inversion) open a bandgap in bulk and monolayer $XTe_2$ (*2*, *4*, *5*). The topological band inversion and the bandgap opening are the hallmarks of a quantum spin Hall state in monolayer 1T' $XTe_2$ (*4-7*). Therefore, a dynamical control of the intralayer Peierls distortion will modulate such exotic quantum electronic properties in $XTe_2$.

Here we report a comprehensive study of the ultrafast structural response in the 1T' and the Td phase of $XTe_2$ (X=Mo, W) with 550 nm and 2000 nm laser excitation. By ultrafast electron diffraction (UED), structure factor calculation and TDDFT-MD simulations, we identify concurrent interlayer and intralayer structural transitions, indicated by the interlayer shear displacement and the reduction of the intralayer Peierls distortion respectively. The interlayer shear mode is well described by a $\pi$-shifted cosine function indicating a displacive excitation of this coherent phonon mode. Therefore, the interlayer structural transition is driven by the electron-phonon coupling, which is in strong contrast to the model of laser filed driven transition in previous study. An ultrafast Mo-Mo (W-W) bond stretching within 0.3 ps is revealed in both the 1T' and the Td phase, which triggers an ultrafast suppression of the intralayer Peierls distortion. We discuss a modification of multiple quantum electronic states associated with the intralayer and interlayer structural transitions. Identifying the complete structural dynamics in atomic and femtosecond spatiotemporal scale pays the way for the ultrafast manipulation of quantum electronic states with photon excitation in $XTe_2$. In addition, the photoinduced concurrent intralayer and interlayer structural transition suggests the traversal of all double-well potential energy surfaces (DWPES), which may be an intrinsic mechanism in the field of symmetry engineering in material system driving by ultrafast photoexcitation.

## Results

### Ultrafast electron diffraction and diffraction pattern of $XTe_2$

The $MoTe_2$ and $WTe_2$ film used in the experiment are prepared by mechanical exfoliation from a bulk crystal (HQ Graphene). The thickness of the freestanding film on the TEM grid is > 30 nm, characterized by the propagation of the breathing phonon mode induced by laser excitation (*26*). A schematic representation of the ultrafast electron diffraction experiment is depicted in Fig. 1C. Visible (550 nm) and midinfrared (2000 nm) femtosecond laser pulses are employed to electronically excite the nanofilm of $MoTe_2$ (and $WTe_2$). We choose these two pump wavelengths in order to separate the possible contributions of the photoexcitation and the electric field to the structural dynamics (*22*, *23*). After laser excitation, another femtosecond electron pulse subsequently probes the structural changes at a varying time delay. The temporal resolution of the system is estimated to be ~ 150 fs (FWHM) (*27*). The phase of $MoTe_2$ in the experiment is controlled by the temperature, i.e. the 1T' phase at room temperature 295 K and the Td phase at 120 K. The $WTe_2$ hold the same phase, i.e. the Td phase, at the room temperature and the low temperature. A prototypical diffraction pattern of the $MoTe_2$ in the Td phase is shown in Fig. 1D.

**Interlayer structural transition by the shear displacement**

To investigate the transient structural dynamics, we focus on the relative intensity change of the Bragg reflections as a function of time delay. For the Td phase of $MoTe_2$, pronounced intensity oscillation of the Bragg reflection is observed with 550 nm and 2000 nm laser excitation, as shown in Fig. 2A and 2B. The frequency of the oscillation is 0.4 THz. Such a low frequency oscillation with a life time of ~ 40 ps is attributed to the interlayer shear phonon mode (*23*, *29*), i.e. the atoms in the same layer vibrate along the same direction while atoms in two adjacent layers vibrate toward opposite directions. The oppositely phased intensity oscillations along the *b* axis (i.e. (h20), (h30), (h40), (h50)) are shown in Fig. 2C and Fig. S1, indicating that the intensity oscillations arises from the interlayer shear phonon (*22*). A similar shear mode oscillation in the sister compound $WTe_2$ is observed and shown in Fig. S2. With the pump fluence increase, the amplitude of the shear mode increases linearly then saturates somewhat at higher fluence (the threshold fluence is 5.44 mJ/cm$^2$ for 550 nm laser pump), as shown in Fig. 2D. A schematic illustration of the interlayer shear displacement is displayed in Fig. 2E. Such an interlayer shear mode has been identified as a signature of an electronic transition from the Weyl semimetal to the trivial phase (*22*, *23*). We simulate the shear displacement induced intensity changes by structure factor calculation (see the details of the calculation in Materials and Methods). In the calculation, to modulate the shear displacement relative to the equilibrium position, the adjacent layers move in opposite direction along *b* axis by 0.015 Å then -0.015 Å. The simulated intensity changes qualitatively agree with the experimental results as shown in Fig. 2F. Except the Td phase, the emergence of unexpected shear mode is observed in the 1T' phase (see Fig. S3). Together with the disappearance of the intensity oscillation for Bragg reflections belonging to the same family of lattice plane (see Fig. S4), we attribute these unexpected structural dynamics to the twin structure and the interlayer stacking fault (see details in section 2 in Supplementary Materials).

In Fig. 2G, the shear phonon mode is fitted by a combination of an exponential function and an exponentially decaying cosine function, i.e. $A*exp(-t/\tau_1)+B*cos(\omega t+\varphi)*exp(-t/\tau_2)$. The frequency $\omega$ and the phase $\varphi$ of the best fit are 0.38±0.001 THz and -0.15±0.01 rad. More fitting and discussion about the fit function are shown in Fig. S5. The identified cosine fitting of the shear mode indicates a displacive excitation of this coherent phonon mode (DECP). In this case, the ultrafast electronic excitation gives rise to the immediate change of the PES (*34*, *35*) and the symmetry changes towards the centrosymmetric phase. A detailed discussion about the symmetry change in $XTe_2$ is shown later in Fig. 5. The timescale of the symmetry change by DECP would be within one quarter of the oscillation period (*37*, *40*). In the case of $XTe_2$, the established new equilibrium position corresponding to the shear mode is achieved within half the period as shown in Fig. 2. The one quarter of the oscillation period (~0.66 ps) is in good agreement with that of the topological switch through ultrafast spectroscopy (~ 0.6 ps in Ref. *23*), evidencing the DECP nature of the topological switch. We do not observe the two-step shear displacement as that in previous study (*54*).

With intensity change at the equilibrium state (≥ 50 ps), we evaluate the interlayer structure transition induced by the shear displacement. Fig. 3A displays the long term evolution of the intensity of the (0k0) reflections in the Td phase of $MoTe_2$ with 550 nm laser excitation. The significant intensity decrease of the (010) reflection than that of other (0k0) reflections suggests the structural response beyond thermal effect. The intensity change of (h00) reflections is used to quantify the Debye Waller effect (i.e. the thermal effect) after laser excitation (see Fig. S6). The corresponding atomic displacement by thermal effect is identified to be ~0.06 Å. In Fig. 3B, introducing the shear displacement gives rise to a larger intensity decrease of the (010) reflection. However, the intensity change of the (020), (040) and (050) reflection still deviates away from the experimental results. Modulating the amplitude of the shear displacement is not working. Note that, besides the interlayer shear displacement, we have identified the intralayer Mo-Mo bond stretching (i.e. the suppression of the Peierls distortion) in the next section, which should be added to the structural response model at equilibrium state. As shown in Fig. 3B, the combination of the Debye Waller (0.06 Å), the shear displacement (0.006 Å) and the Mo-Mo bond stretching (0.006 Å) gives rise to a good agreement between the calculated intensity change and the experiment results (see detail on the fit in section 5 in Supplementary Materials). Based on the same model, the good agreement between calculation and experiment for the (1k0) reflection is shown in Fig. 3C and 3D. The same experimental structure response is also observed in $MoTe_2$ with 2000 nm laser excitation and $WTe_2$ (see Fig. S7). Therefore, we identify both the interlayer and intralayer structural transition at the equilibrium state ($\geq$ 50 ps).

**Intralayer structural transition by the suppression of the Peierls distortion**

To gain further insight into the ultrafast structural response except the shear mode, we focus on the intensity change of Bragg reflections on femtosecond timescale. For the 1T' phase, the time-dependent intensity change is shown in Fig. 4A with 2000 nm laser excitation. Within $t_1$ = 0.3 ps, anisotrpic intensity changes for Bragg reflections are observed, for example, the intensity of the (040), (120), and (320) reflection decays significantly, while the intensity of the (060), (130) and (330) reflection stays unchanged. The time constant for the intensity decay of (120) is 132±37 fs fitted by an exponential function, as shown in the inset in Fig. 4A, which is much faster than the equilibrium of the overall lattice system (see Fig. S6D). The same anisotropic intensity change is observed in the Td phase of $MoTe_2$ (as shown in Fig. 4E) and $WTe_2$ (see Fig. S2). Since both the 1T' and Td phase exhibit the same intralayer Peierls distortion, we speculate a possible structural transition associated with the suppression of the intralayer Peierls distortion (*2*). The anisotropic intensity change within 0.3 ps is against the trend of the intensity modulation by the shear displacement, as shown in Fig. 4E, so the dominating structural response is not the interlayer transition in this time scale.

For $XTe_2$, the Fermi surface nesting drives the intralayer Peierls distortion (*2*), characterized by the in-plane Mo-Mo metallic bonds and the out-of-plane wrinkling of Te and Mo atoms, as shown in Fig. 1A. Generally, the femtosecond laser excitation induces the flattening of the double-well potential energy surface and subsequently suppresses the structural distortion to a higher crystal symmetry in many material systems (*36-41*). In the case of $MoTe_2$, the femtosecond laser induces a population in the antibonding d-orbitals of Mo atoms (*42*, *43*), then the shorter Mo-Mo distance (i.e. the Mo-Mo metallic bonds) could get elongated. By structure factor calculation, we calculate the intensity change of Bragg reflections by introducing a Mo-Mo bond stretching in the unit cell of the 1T' phase and the Td phase. The possible reduction of the out-of-plane wrinkling along the c axis is neglected in the calculation, since the experiment is not sensitive to such out-of-plane motions in this geometry. Fig. 4B and Fig. 4F display the calculated intensity change as a function of the Mo-Mo bond stretching in the 1T' phase and the Td phase of $MoTe_2$. The similar anisotropic intensity changes for Bragg reflections as that of experiment results in Fig. 4A and 4E are observed in the calculation results. The bar chart in Fig. 4C and Fig. 4G shows the qualitative agreement between the calculation and the experiment results. The discrepancy between the quantified intensity change of the experiment and the calculation results in the bar chart may derive from the neglected Debye Waller of Mo and Te, the concurrent shear displacement and the interlayer stacking fault in the sample. The same structural response is observed with both 550 nm and 2000 nm laser excitation in the Td phase of $MoTe_2$ (see Fig. S8). By structure factor calculation of the element-dependent Debye Waller effect (see Fig. S9), we conclude that the Debye Waller effect plays a minor role in the structural response within 0.3 ps. Therefore, together with the structural response at the equilibrium state (≥ 50 ps) in Fig. 3, we attribute the ultrafast anisotropic intensity changes to the intralayer structure transition. i.e. the Mo-Mo (W-W) bond stretching, in $XTe_2$. The Ag mode with a frequency of 112 $cm^{-1}$ could be the phonon mode dominating the intralayer structure transition (*53*).

**Photoinduced Intralayer and interlayer atomic motions from TDDFT-MD simulation**

To further confirm the concurrent intralayer and interlayer atomic motions, we perform TDDFT-MD simulations for $MoTe_2$ in its Td phase. The pump fluence is set to 2 $mJ/cm^2$ close to the experimental condition. More detailed information regarding the simulation can be found in Supplementary Materials. The simulation results are shown in Fig. 5. The arrows in Fig. 5A indicate the averaged directional movements of Mo and Te atoms in the unit cell. Detailed displacement trajectories with time for atoms in the middle layer (the bottom layer) are summarized in Fig. 5B and 5C (and Fig. S10B to S10D). As shown in Fig. 5B (bottom), the bond length of Mo1 and Mo2 stretches significantly within ~ 0.3 ps, which agrees with the ultrafast bond stretching from the experimental results. Meanwhile, Te8 and Te11 move in opposite directions along the c axis as shown in Fig. 5C (top), which reduces the out-of-plane wrinkling. The stretching of the metallic bonds and the reduction of the out-of-plane wrinkling within 0.3 ps evidence the photoinduced suppression of the intralayer Peierls distortion. Note that T8 and Te11 also move simultaneously along the negative direction of *b* axis as shown in Fig. 5C (bottom), while Te7 and Te12 in adjacent layer move along the positive direction of the *b* axis (see Fig. S10C (bottom)). Such opposite movement along the *b* axis in two adjacent layers is a signature of the interlayer shear displacement. The intensity change of several Bragg reflections is calculated based on the atomic displacements at 0.2 ps in Fig. 5. The calculated anisotropic intensity change, as shown in Fig. S11, agrees qualitatively with the experiment results in Fig. 4. The simulation results in Fig. 5 contains some coherent oscillations of the atomic displacements with the period of 200-300 fs, which are not observed in experimental results. Such coherent oscillations could be attributed to the excitation of high-frequency (~ 4 THz) phonon modes (*24*, *31*). The ~150 fs temporal resolution of our experimental system is insufficient to detect such ultrafast oscillation.

Overall, the simulation results in Fig. 5 unambiguously demonstrate the concurrent interlayer shear displacement and the suppression of the intralayer Peierls distortion, in agreement with the experiment results. A schematic illustration of such a structural change is summarized in Fig. 5A. On sub-ps timescales, the intralayer Peierls distortion is suppressed to a 1T-like structure. Meanwhile, the shear displacement reduces the bond length discrepancy between $d_1$ and $d_3$ and the interlayer stacking is changed correspondingly as shown in Fig. 5A. When the symmetry center of the top (bottom) layer (i.e. the gray crosses) gets aligned to that of the center layer (i.e. the red cross), a structural transition from the non-cenrosymmetric to the centroymmetric can be achieved by the shear displacement. We define an intermediate centrosymmetric 1T(*) phase with the 1T-like intralayer structure (Fig. 5A (right)). Therefore, the photoexcitation induces a structural transformation from the non-centersymmetric Td phase (Fig. 5A (left)) to the centrosymmetric 1T(*) phase (Fig. 5A (right)). Note that the unit cell of the 1T(*) phase, indicated by the dotted rectangular in Fig. 5A (right), will be half of the unit cell of the Td phase (the gray rectangle) along the *b* axis, which may be a signature for further experimental study on such a structural transition. In our experiment, the inhomogeneous longitudinal excitation due to the limited optical penetration depth and the interlayer stacking fault may blur the underlying unit cell change.

## Discussion

In this work, we reveal the photoexcitation induced concurrent intralayer and interlayer structural transitions in the 1T' and the Td phase of $XTe_2$ (X=Mo, W) by femtosecond electron diffraction, structure factor calculation and TDDFT-MD simulations. The pathway of the concurrent structural transitions in real space and the sketch of the modulation of the PES by photoexcitation for the Td phase of $MoTe_2$ are illustrated in Fig. 6. After photoexcitation, the occupation of the antibonding d-orbitals of Mo atoms induces the stretching of the in-plane metallic Mo-Mo bonds. It is conceivable that at higher pump fluence, the Mo-Mo bond will dissociate completely. The Mo-Mo bonds stretching and the reduction of the out-of-plane wrinkling of Te atoms give rise to an ultrafast suppression of the Peierls distortion and a transition to intralayer 1T-like structure within 0.3 ps. Meanwhile, the photoinduced interlayer shear displacement produces a transition to a centrosymmetric phase in sub-period of the shear phonon mode. The multidimensional potential energy surfaces at the ground state in the Td phase of $XTe_2$ is shown in Fig. 6B (top). The photoexcitation flattens the DWPES along the Peierls distortion coordinate and the shear mode coordinate, as shown in Fig. 6B (bottom), giving rise to the concurrent intralayer and interlayer structure transition. A simple question is that if there is any correlation between the intralayer and interlayer structural transitions, for example, the intralayer structural transition will facilitate or impede the interlayer shear transition, which needs to be studied further. In contrast to the complete interlayer transition on the surface layer of bulk $MoTe_2$ by ultrafast spectroscopy (*23*), the displacement corresponding to both the interlayer and intralayer transition is ~0.01Å in our UED experiment, much smaller than the required displacement for a complete transition (0.19 Å and 0.43 Å respectively). The discrepancy may derive from the limited optical penetration depth and the interlayer stacking fault. The twin structure and the stacking fault in both the 1T' and the Td phase of $XTe_2$ are identified by the ultrafast structural response. Ultrafast electron diffraction is expected to be a powerful tool to reveal twin structures and stacking faults in broad material system.

The significance of the revealed ultrafast structural response in $XTe_2$ (X=Mo, W) in our work is multifold. First, we provide a deep insight into the ultrafast structural response in atomic and femtosecond spatiotemporal scale. We clearly identify the ultrafast intralayer structural transition and the displacive excitation of the interlayer shear mode. Therefore, the electron-phonon coupling is demonstrated to be the driving force of a photoinduced Weyl semimetal phase to trivial phase transition, which is distinct from the model of THz/light field driven transition in Ref. *22-24*; Second, the intralayer Peierls distortion causes the band inversion (*2*, *17*, *14*), consequently, the topological insulator phase in monolayer (*2*, *4*, *5*, *17*) and the higher-order topological insulator phase in the bulk 1T' and Td phase (*13*, *14*, *16*) are expected to switch to the trivial phase upon suppression of the intralayer Peierls distortion. Moreover, the quantum spin Hall state based on the topological band inversion in the monolayer 1T' phase of $XTe_2$ (*4-7*) will also be modulated by the suppression of the intralayer Peierls distortion; Third, the ultrafast suppression of the intralayer Peierls distortion can be used to tune the anisotropic electronic and optical properties in the 1T' phase of TMDCs (*25*) for practical applications.

The recent study suggests an ultrafast Lifshitz transition within 0.4 ps in the Td phase of $MoTe_2$ (*44*). Such an electronic transition is attributed to the dynamical modification of the Coulomb interaction, with the absence of precise knowledge on the structural response in the sub-ps timescale. The ultrafast suppression of the intralayer Peierls distortion within 0.3 ps, revealed in our work, will modify the Fermi surface and therefore may contribute to the ultrafast Lifshitz transition. Further study is required to identify the detailed correlation between the structural response and the electronic transition.

The photoexcitation-driven modification of the potential energy surface and the associated structural transition present a new opportunity for manipulating the properties of materials and have been widely explored in material system (*36-41*), but in these study, the structural transition mainly contains or solely discuss single DWPES (which is also known as potential energy surface with saddle point) and the order parameter on a simple potential energy surface. In our results of $XTe_2$ (X=Mo, W), the concurrent interlayer and intralayer structural transitions indicate the traversal of the two DWPES by photoexcitation, which is beyond the previous concept for the single DWPES based structural transition. Consequently, the traversal of all DWPES achievable by the photon-generated carriers is conceivable in complex material system with multidimensional potential energy surfaces (i.e. multidimensional structural distortions). In other words, the traversal of all DWPES may be the intrinsic mechanism dominating the photoexcitation-driven structural transition. The recent study suggests the disorder (or multi-mode) driven ultrafast structural transition and the order parameter on a simple potential energy surface plays a minor role (*38*, *45*). In our work, the concept of the traversal of all DWPES in photoexcitation-driven structural transition, involving coherent and incoherent evolution of multi phonon modes, may be a clearer transition regime than the disorder-driven transition model (*38*, *45*). Except the above discussed global structure transition, photo-induced ultrafast transition of the local structural distortion associated with the local DWPES, is also reported recently (*46*). The traversal of all DWPES in material system can be used to guide the property control by photoexcitation-driven symmetry engineering and also be used to identify the multidimensional local energy minimum. Ultrafast electron/X-ray diffraction is a powerful tool leading the research along this way. Since structure determines property is the traditional paradigm of materials science, the traversal of DWPES and the associated structural transition suggest determining the ultrafast structural response before attributing photo-induced exotic property change to the perturbation of other degrees of freedom (such as electron and spin) in material system.

## Materials and Methods

## Ultrafast electron diffraction experiment

In the ultrafast electron diffraction experiment, a 70-keV DC electron diffraction setup is used to study the ultrafast atomic motion during the photoinduced phase transitions. The electron pulses were set to a few thousand electrons per pulse with a spot size of ~100 μm diameter at the sample position. The transverse spot size of the pump laser is ~400 μm, which is much larger than the spot size of the electron pulse to keep a relative homogeneous excitation of the detected area on the sample. The repetition rate of the pump laser is 1 kHz and the the pump wavelength is tuned by commercial TOPAS facility. The instrument response function duration of the UED system is estimated to be ~ 150 fs FWHM (*27*). The base temperature of the sample is controlled by liquid nitrogen. The thickness of the freestanding film on the TEM grid is characterized by the propagation of the breathing phonon mode induced by laser excitation. Though a weak breathing mode is measured suggesting the thickness of ~ 30 nm, the exfoliated film is not that homogeneous indicating the overall thickness of the film could be larger than 30 nm. The structural dynamics measurements in the text also suggest the thickness of the film > 30 nm.

## Structure factor calculation for the interlayer and intralayer structure transition and the Debye Waller effect.

The intensity of a Bragg peak, $I \propto |F|^2$, can be calculated using the structure factor:

$$F(hkl) = \sum_j f_j exp\,[-i2\pi(hx_j + ky_j + lz_j)] \quad (1)$$

Where the summation runs over all atoms in the unit cell (four Mo and eight Te atoms), $f_j$ is the atomic scattering factor for the jth atom, $r_j = x_j\hat{a} + y_j\hat{b} + z_j\hat{c}$ is the vector position of the atom in the unit cell and *(hkl)* is the Miller indices. For the Debye Waller effect, we introduce the atomic Debye Waller factor into the structure factor calculation:

$$T_j = \exp(-M), \; M = 8\pi^2\langle\mu^2\rangle(sin\theta/\lambda)^2 \quad (2)$$

$M$ denotes the Debye Waller factor and $\langle\mu^2\rangle$ represents the mean square

displacement of Mo or Te atom.

To simulate the interlayer and intralayer structure transition, we calculate the intensity modulation $\Delta I/I_0$ by introducing the interlayer shear displacement and the Mo-Mo bond stretching. For the interlayer shear displacement of the trilayer construction in the unit cell, we define the atoms in the top and bottom layer displace $\Delta y$ while the atoms in the medium layer displace $-\Delta y$. The structure factor in this case is:

$$F(hkl, \Delta y) = \sum_{top\ and\ bottom} f_j T_j exp\,[-i2\pi(hx_j + k(y_j + \Delta y) + lz_j)] + \sum_{medium} f_j T_j exp\,[-i2\pi(hx_j + k(y_j - \Delta y) + lz_j)] \quad (3)$$

For the intralayer Mo-Mo bond stretching, the displacement of Mo atoms is introduced into the above equation in a similar way as that of the shear displacement. The structure factor calculated by the algebra equation has been compared with that from SingleCrystal to confirm reliable structure factor calculation in this work.

## Methods for TDDFT-MD simulations

The experimental geometry of the bulk $MoTe_2$ is adopted, which is characterized by an orthorhombic ($T_d$) unit cell without inversion symmetry (*47*). To study the optoelectronic responses of $MoTe_2$ in $T_d$ phase, linearly polarized laser beams with time-dependent electric field $E(t) = E_0 \cos(\omega t) \exp[-(t-t_0)^2 / 2\sigma^2]$ are applied along the crystallographic a-axis. The photon energy, width, and amplitude are set as 2.25 eV (λ = 550 nm), 15 fs and 0.06 V/Å, respectively. This setup allows us to reproduce a laser fluence (ca. 2 mJ/cm$^2$) similar to experimental measurements.

The TDDFT-MD calculations are performed using the time dependent ab initio package (TDAP) as implemented in SIESTA (*48-50*). The bulk $MoTe_2$ in its $T_d$ phase is simulated with a unit cell of 12 atoms with periodical boundary conditions. Numerical atomic orbitals with double zeta polarization (DZP) are employed as the basis set. The electron-nuclear interactions are described by Troullier-Martins pseudopotentials, PBE functional (*51*). An auxiliary real-space grid equivalent to a plane-wave cutoff of 250 Ry is adopted. To make a good balance between the calculation precision and cost, a Γ-centered 6 × 5 × 3 k-point grid is used to sample the Brillouin zone. The coupling between atomic and electronic motions is governed by the Ehrenfest approximation (*52*). During dynamic simulations the evolving time step is set to 0.05 fs for both electrons and ions in a micro-canonical ensemble.

## Acknowledgements

**Funding:** This work was funded by the Max Planck Society and the European Research Council (ERC) under the European Union's Horizon 2020 research and innovation program (Grant Agreement Number ERC-2015-CoG-682843). Yingpeng Qi acknowledges support by the Sino-German (CSC-DAAD) Postdoc Scholarship Program (Grant No. 201709920054 and No. 57343410) and the funding from Max Plank Society. **Author Contributions:** Y. Qi and R. Ernstorfer designed the experiments. Y. Qi, D. Zahn, T. Vasileiadis and H. Seiler executed the experiments. Y. Qi did the data analysis and the structure factor calculation. M. Guan, H. Zhao and S. Meng did the TDDFT-MD simulation. Y. Qi wrote the manuscript with contributions from all the authors. **Competing interests:** The authors declare no competing interest. **Data and materials availability:** All data needed to evaluate the conclusions in the paper are present in the paper and/or the Supplementary Materials. Additional data related to this paper may be requested from the authors.

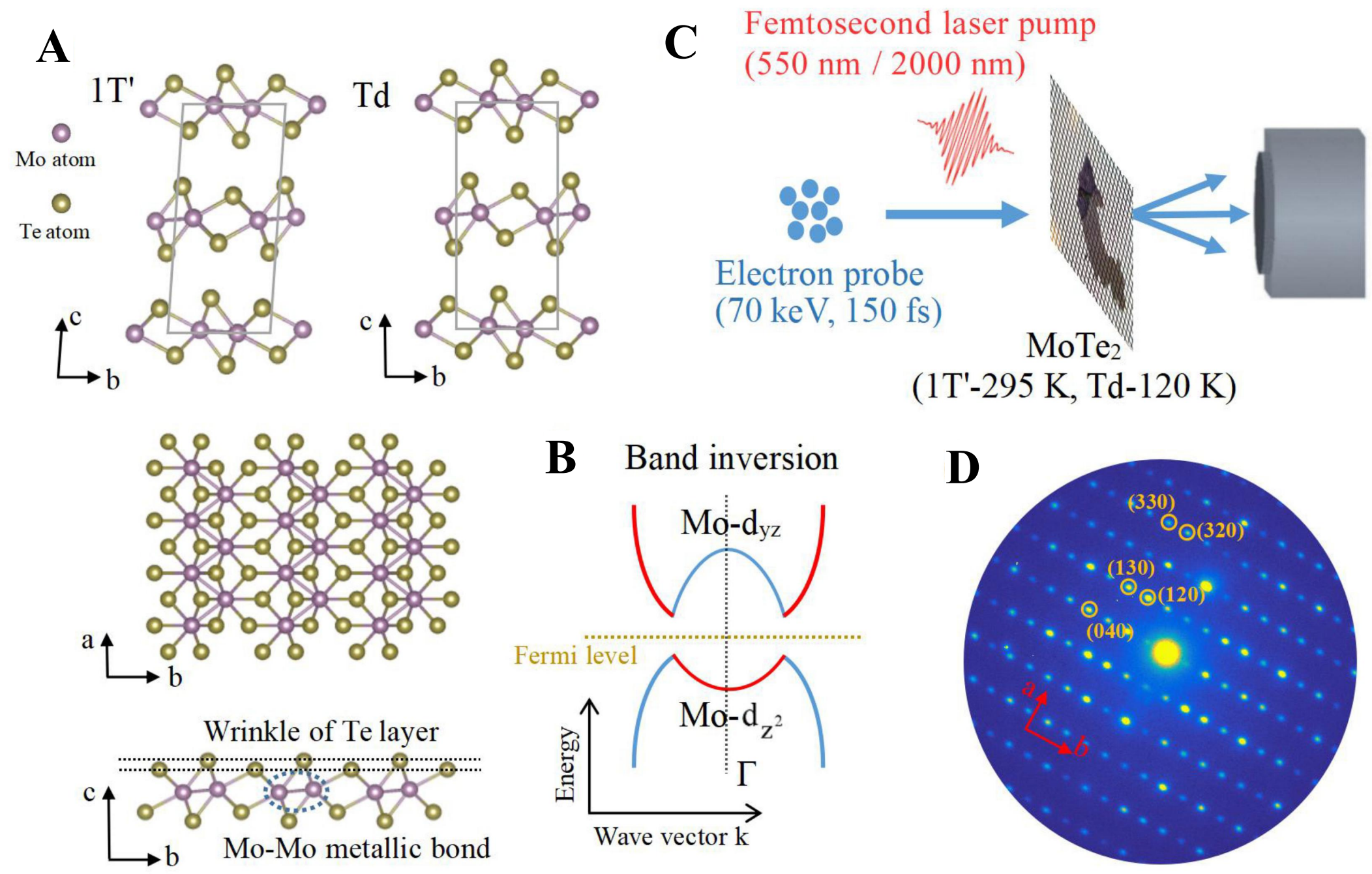

**Fig. 1. Crystal structure of $MoTe_2$ and the diffraction pattern obtained in femtosecond electron diffraction experiment.** (**A**) The unit cell of $MoTe_2$ in the monoclinic 1T' phase and the orthorhombic Td phase. The bottom shows the top view (a-b plane) and the side view (b-c plane) of the crystal structure of a single layer. The two phases hold the same in-plane crystal structure but different vertical stacking. All the drawings of the crystal structure are produced by VESTA software (*28*). (**B**) Schematic band inversion by intralayer Peierls distortion in monolayer 1T' $MoTe_2$. (**C**) Schematic presentation of the ultrafast electron diffraction experiment. The femtosecond laser pumps the $MoTe_2$ nanofilm with the crystal phase controlled by the temperature. Another femtosecond electron pulse diffracts off the crystal, thus probing transient structural changes. (**D**) A prototypical diffraction pattern of $MoTe_2$ in the experiment. Several spots are labeled by circles.

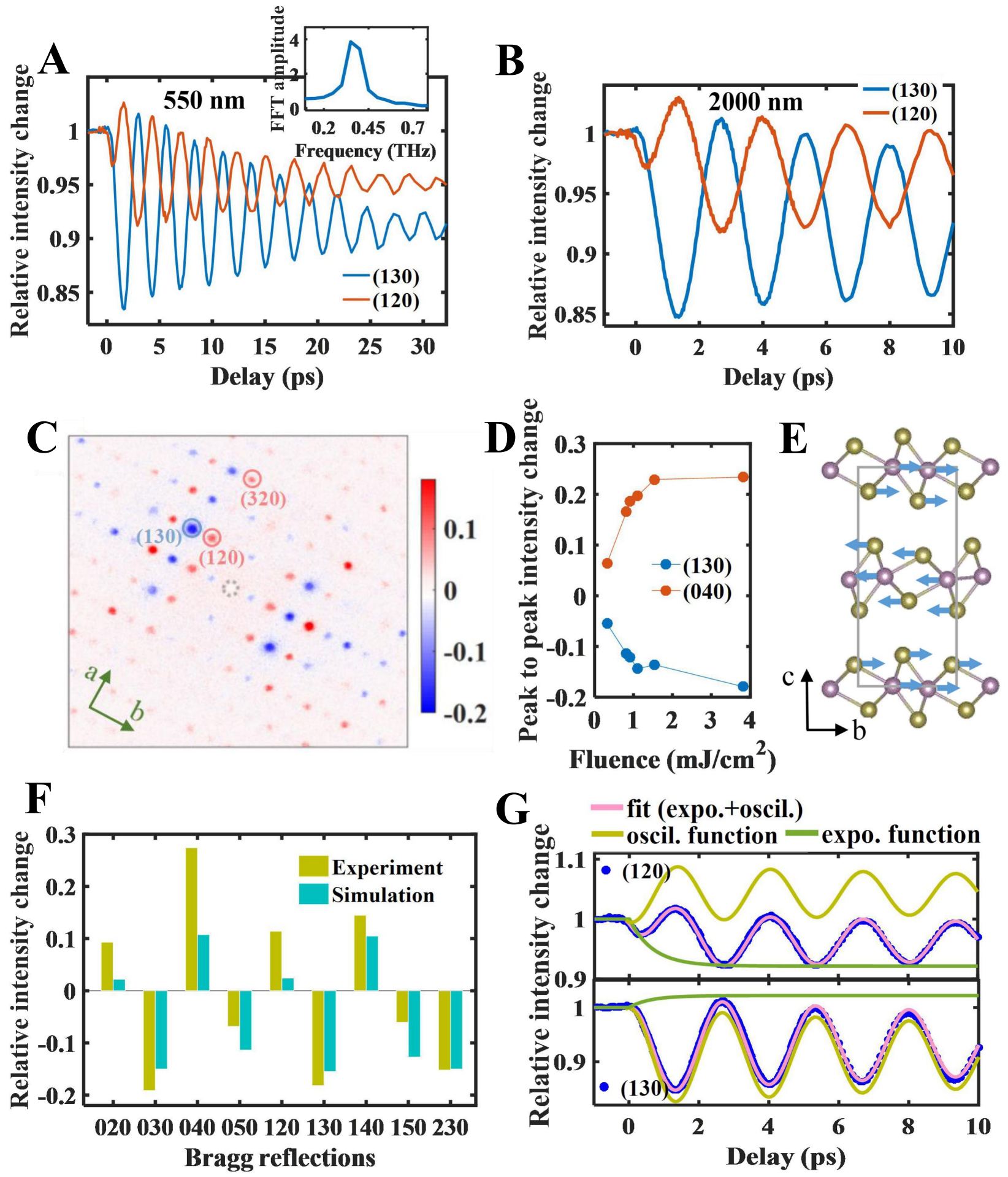

**Fig. 2. Photoinduced interlayer shear mode in the Td phase of $MoTe_2$.** (**A**) and (**B**) are intensity changes with 550 nm (3.81 mJ/cm$^2$) and 2000 nm (15.08 mJ/cm$^2$) laser pump. The Fast Fourier Transformation (FFT) amplitude of oscillations in the inset indicates the ~0.4 THz shear phonon mode. (**C**) Difference map of the intensity between the delay points of 1.6 ps and 2.9 ps with 550 (3.81 mJ/cm$^2$) nm laser pump. (**D**) The peak-to-peak intensity change of the intensity oscillation of (130) and (040) reflection as a function of the pump fluence. (**E**) Schematic illustration of the interlayer shear mode. (**F**) Bar chart showing the shear mode induced intensity changes of several peaks in experiment (550 nm and 3.81 mJ/cm$^2$) and simulation. The shear displacement of each layer is 0.015 Å. (**G**) The fit of the intensity oscillation of the (120) and (130) reflection. The pink solid curve is the best fit, which is composed of an exponential function (the green curve) and an exponentially decaying cosine function (the yellow curve). The pump laser is 2000 nm and 15.08 mJ/cm$^2$.

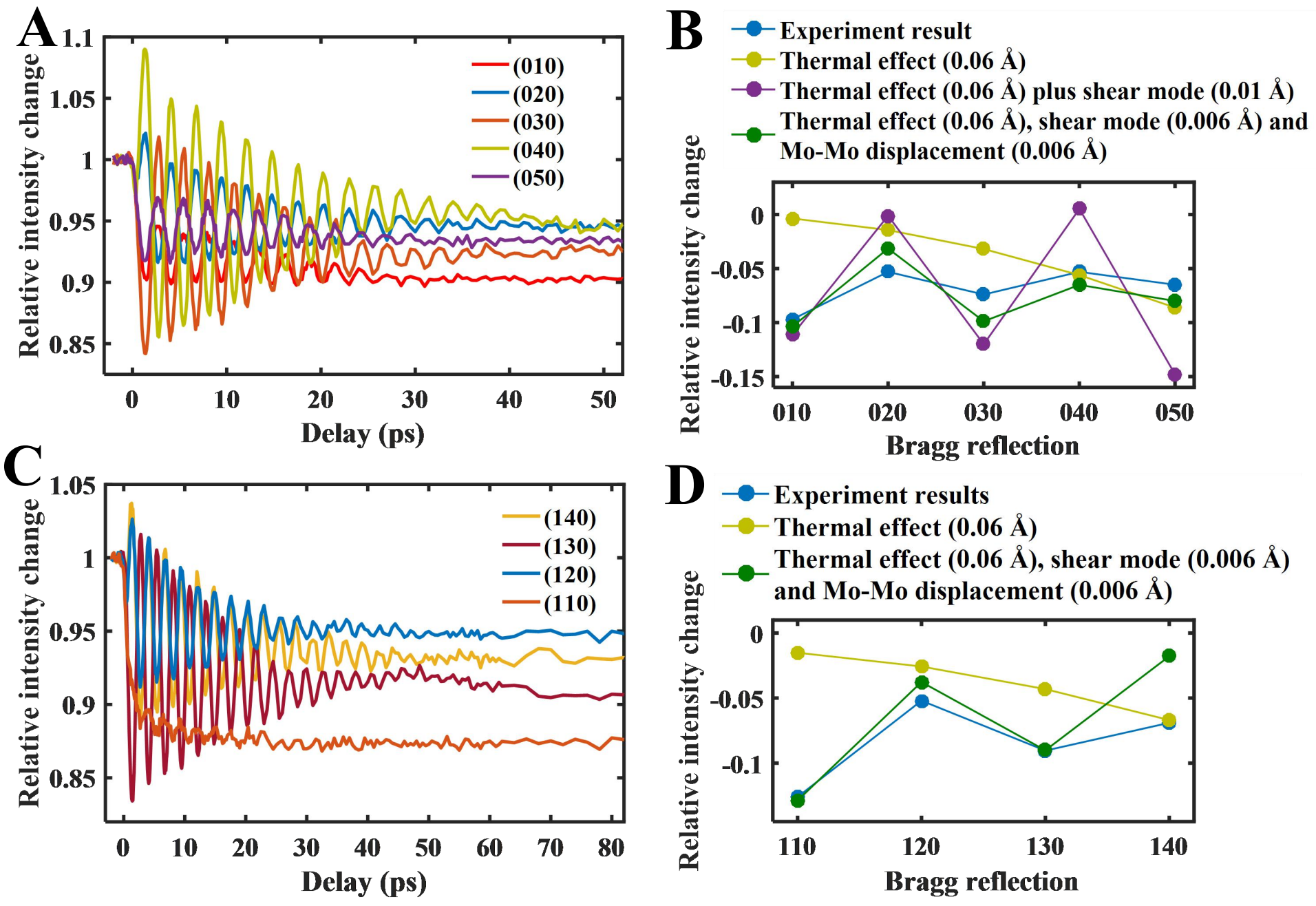

**Fig. 3. Photoinduced interlayer structure transition by the shear displacement in the Td phase of $MoTe_2$.** The pump laser is 550 nm and 3.81 mJ/cm$^2$. (**A**) The long term evolution of the intensity of the (0k0) reflections. (**B**) The experimental intensity change of (0k0) at the time delay of 50 ps and the structure factor calculation of the intensity change. The thermal effect, the interlayer shear displacement and the intralayer Mo-Mo bond stretching are involved in the calculation. (**C**) The long term evolution of the intensity of the (1k0) reflections. (**D**) The experimental intensity change of (1k0) at the time delay of 80 ps and the structure factor calculation of the intensity change.

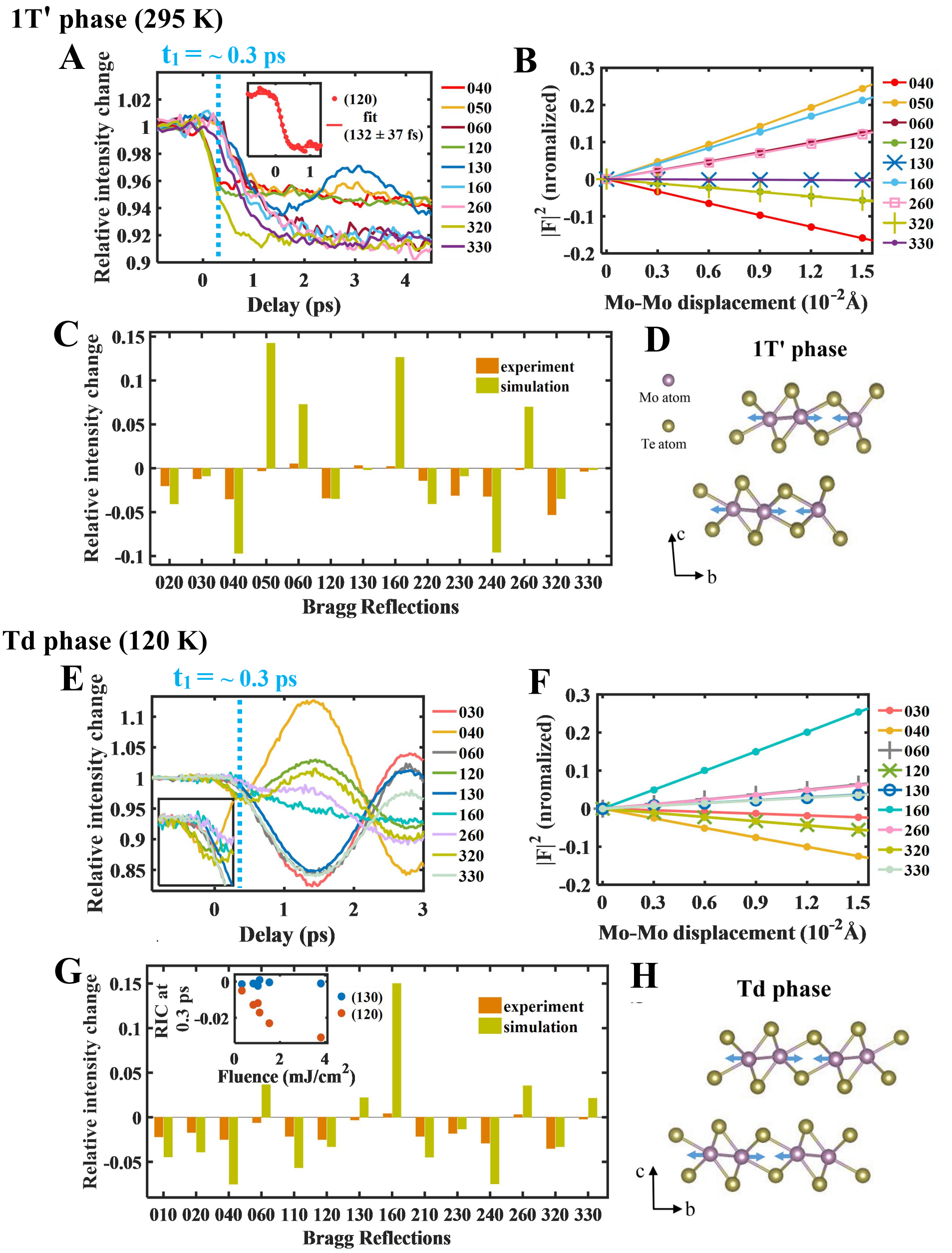

**Fig. 4. Photoinduced intralayer structure transition by Mo-Mo bond stretching in the 1T' (A-D) and Td (E-H) phase of $MoTe_2$.** With 2000 nm (13.10 mJ/cm$^2$) laser excitation, (**A**) anisotropic intensity change (a prompt decay vs a delayed decay) of Bragg reflections in the 1T' phase within ~ 0.3 ps. The inset is the fit of the intensity decay of the (120) reflection. (**B**) Structure factor calculation of the intensity changes as a function of Mo-Mo displacement (bond stretching). (**C**) Bar chart showing the Calculated intensity changes vs the experimental results. The Mo-Mo displacement is 0.006 Å (bond stretching of 0.012 Å). (**D**) (left) Schematic illustration of the Mo-Mo bond stretching in the 1T' phase. With 2000 nm (15.08 mJ/cm$^2$) laser excitation, (**E**) anisotropic intensity change of Bragg reflections in the Td phase within ~ 0.3 ps. The inset zoom in the intensity change in sub-ps. (**F**) Structure factor calculation of the intensity changes as a function of Mo-Mo displacement. (**G**) Bar chart showing the Calculated intensity changes vs the experimental results. The Mo-Mo displacement is 0.006 Å. The inset is, with 550 nm laser excitation, the fluence dependence of the intensity changes of (120) and (130) reflection at the time delay of 0.3 ps. (**H**) Schematic illustration of the Mo-Mo bond stretching in theTd phase.

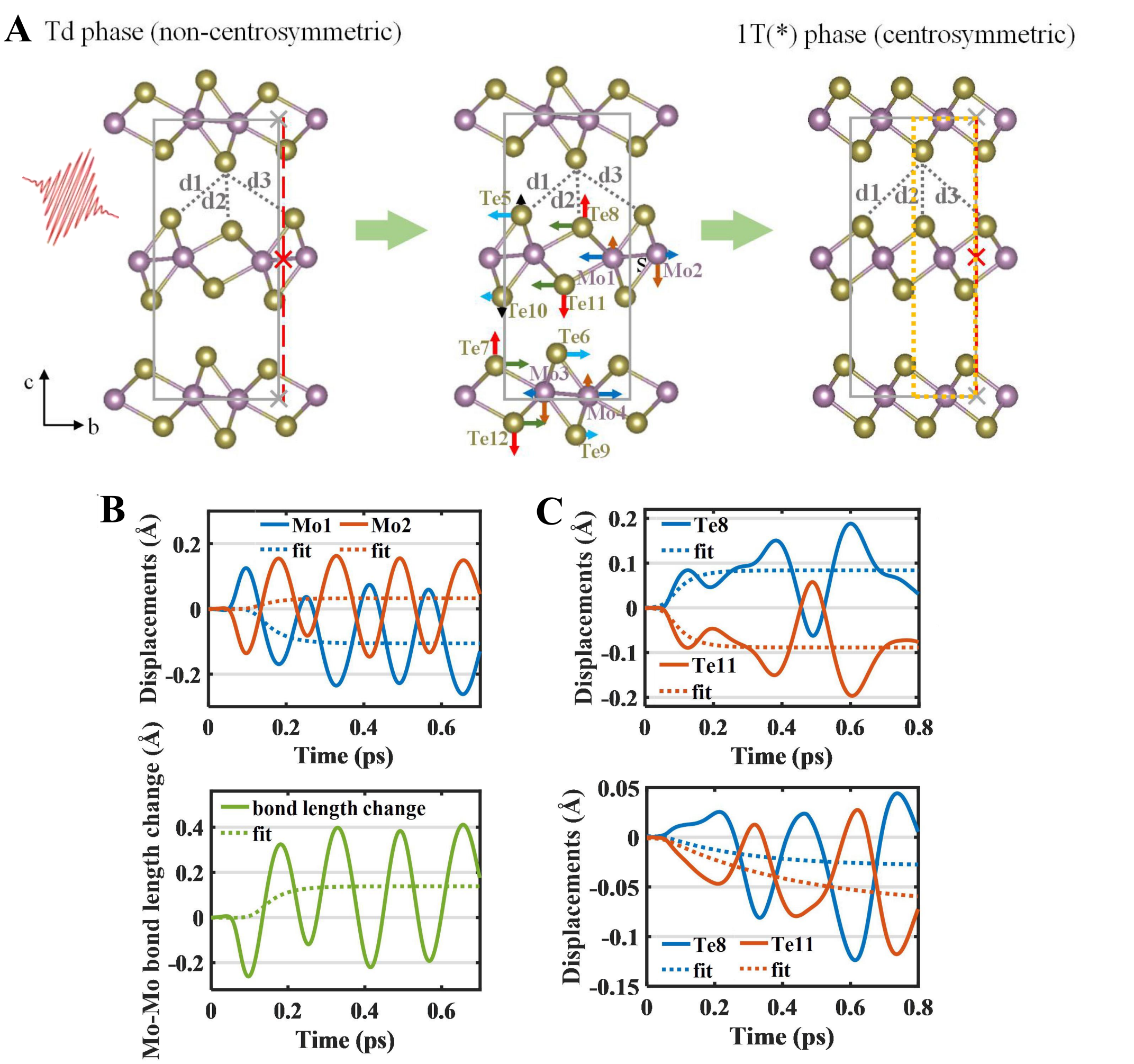

**Fig. 5. Photoinduced intralayer and interlayer structural transition in the Td phase of $MoTe_2$ by TDDFT-MD simulations.** (**A**) In-plane and out-of-plane movements of Mo and Te atoms indicated by arrows. The gray rectangle is the unit cell of the Td phase. The gray crosses are symmetry centers of top and bottom layers and the red cross is the symmetry center of the middle layer. An intermediate state 1T(*) is formed with an intralayer structure akin to the undistorted 1T phase. The bond length $d_3 > d_1$ in the Td phase, while $d_3 = d_1$ in the 1T(*) phase. The dotted yellow rectangular indicates the unit cell of the 1T(*) phase. (**B**) (top) The time-dependent displacements of Mo1 and Mo2 along the b axis. (bottom) The time-dependent Mo-Mo bond length change between Mo1 and Mo2. The dotted curves are the monoexponential fit of the simulation results. (**C**) The time-dependent displacements of Te8 and Te11 along the c axis (top) and the b axis (bottom). The dotted curves are the monoexponential fit of the simulation results.

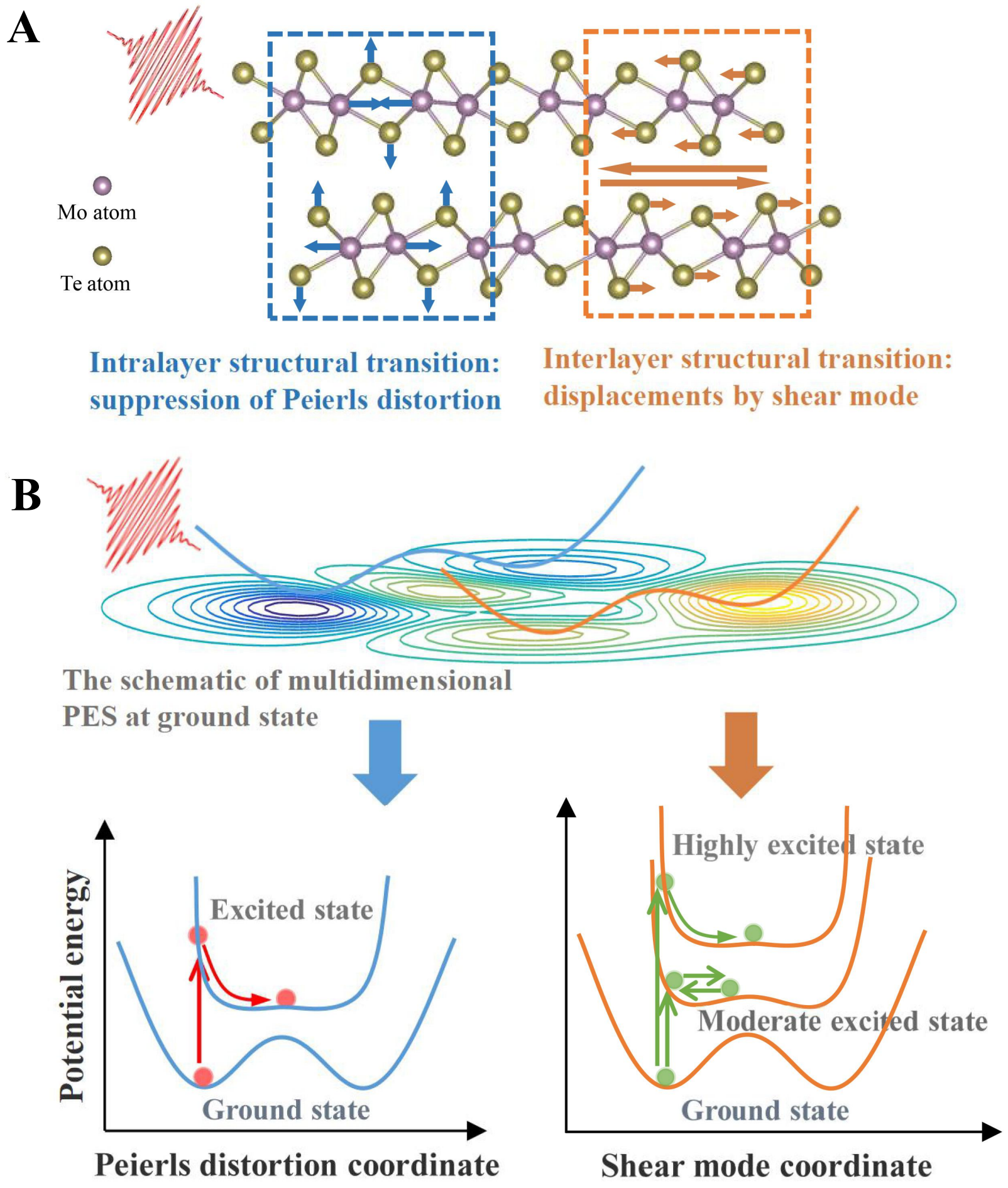

**Fig. 6. Schematic presentation of the concurrent intralayer and interlayer structural transitions in real space and the corresponding changes of the double-well potential energy surfaces in the Td phase of $XTe_2$.** (**A**) Photoinduced intralayer structural transition by suppression of the Peierls distortion and interlayer structural transition by the shear displacement. The main atomic motions in the intralayer transition include the dissociation of Mo-Mo bonds and the reduction of the out-of-plane wrinkling of Te atoms. (**B**) (top) The Schematic illustration of the multidimensional potential energy surfaces at the ground state in the Td phase of $XTe_2$. (bottom) The schematic of the modulation of the potential energy surface of the Peierls distortion and the shear mode by photoexcitation. The flattened potential energy surface gives rise to the suppression of Peierls transition to an 1T-like intralayer structure. For the potential energy surface of the shear mode, at low excitation density, a coherent shear phonon mode is excited. At high excitation density, the symmetry switches from noncentrosymmetry to centrosymmetry in a sub-period of the shear phonon mode.

# Supplemental Materials

1. Photoinduced shear mode in $XTe_2$
2. Stacking fault induced emergence and disappearance of the intensity oscillation in the 1T' and Td phase of $XTe_2$
3. The fit of the shear mode induced intensity oscillation
4. Evaluation of the Debye Waller effect at the equilibrium state
5. Shear displacement induced interlayer structural transition in the Td phase of $MoTe_2$ and $WTe_2$
6. Element-dependent Debye Waller effect to the structural response within 0.3 ps
7. The same structural response with 550 nm and 2000 nm laser excitation for the Td phase of MoTe2
8. TDDFT-MD simulation of the interlayer and intralayer structure transitions

## 1. Photoinduced shear mode in $XTe_2$

The interlayer shear mode induced intensity oscillation of Bragg reflection is identified in both the Td phase of $MoTe_2$ and $WTe_2$, as shown in Fig. S1 and S2. Moreover, Fig. S1 and S2 show that both 550 nm and 2000 nm laser excitation can give rise to the interlayer shear mode.

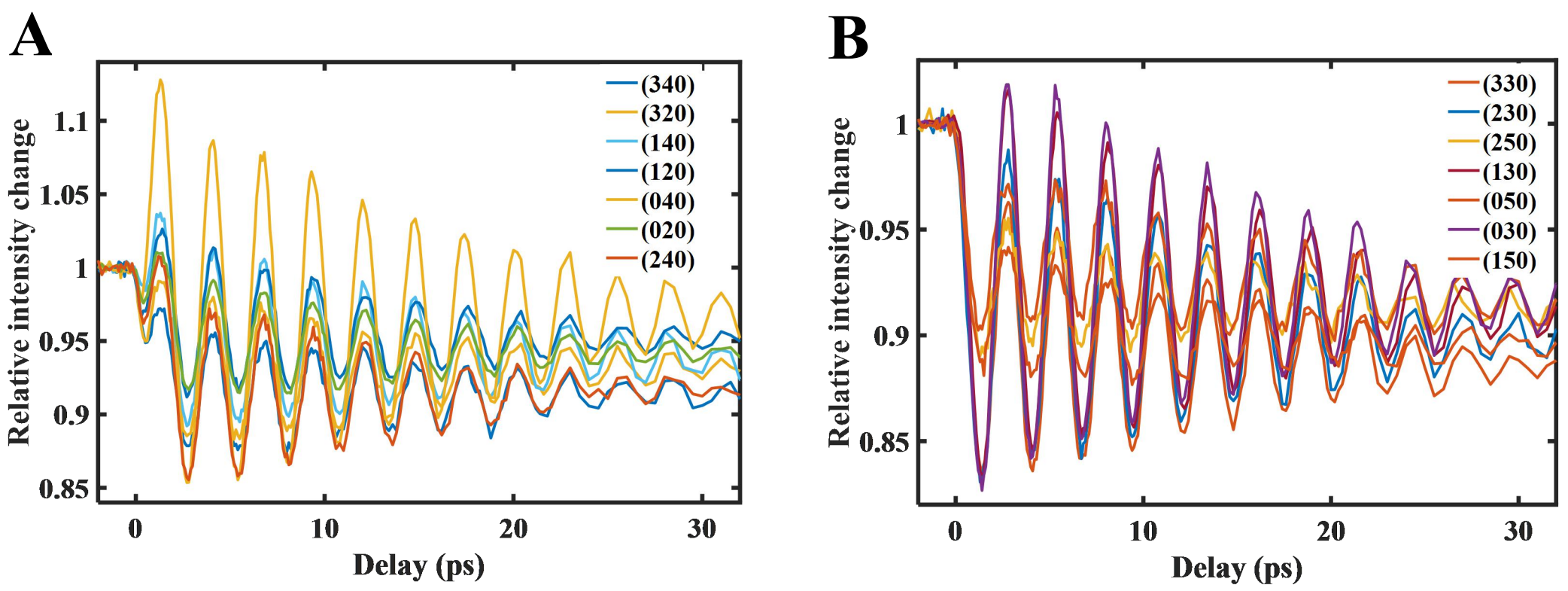

Fig. S1. (A), (B) Intensity oscillation of Bragg spots modulated by the shear phonon mode in the Td phase of $MoTe_2$. The pump laser is 550 nm and 3.81 $mJ/cm^2$. The intensities of (h20) and (h40) in (c) oscillate in the opposite phase of (h30) and (h50) in (d). The alternate, oppositely phased intensity oscillation along the *b* axis of the unit cell indicates the intensity modulation arises from the interlayer shear phonon.

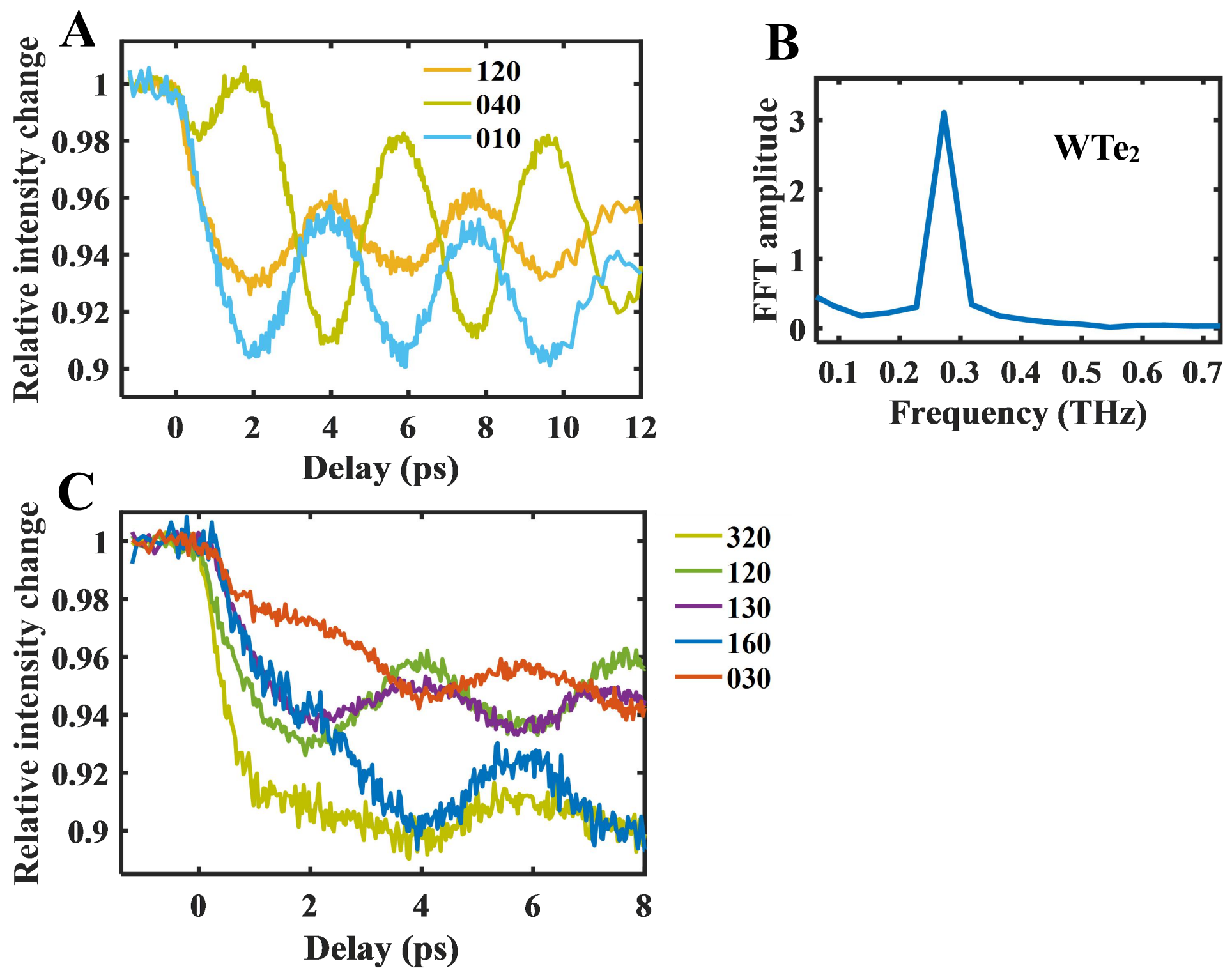

Fig. S2. Time-resolved intensity changes of Bragg reflections in $WTe_2$ with 2000 nm laser excitation (13.90 $mJ/cm^2$) at 110 K. (A) Intensity oscillation of Bragg spots modulated by the shear phonon mode. (B) The FFT amplitude of the oscillation indicates a frequency of 0.27 THz, which is close to the eigenmode of the shear phonon in the Td phase of $WTe_2$. (C) Time-delayed intensity decay of the (030), (130), (160) in contrast to the instantaneous intensity decay of the (120) and (320) after photoexcitation.

## 2. Stacking fault induced emergence and disappearance of the intensity oscillation in the 1T' and Td phase of $XTe_2$

In the 1T' phase of $MoTe_2$, intensity oscillations of the (130) and (120) Bragg reflections are observed with 550 nm and 2000 nm laser excitation, as shown in Fig.

S3A and B. The Fast Fourier Transformation (FFT) amplitude of the oscillation is 0.38 THz, in good agreement with the frequency of the interlayer shear phonon in the Td phase (see Fig. 2 in the main text). Notice that the shear phonon mode is Raman inactive in the 1T' phase (*1*, *2*).The recent spectroscopic study also reports the ermergence of the photoinduced shear phonon mode in the 1T' phase of $MoTe_2$ (*3*). The ermergence of Raman-inactive phonon mode is generally attributed to the breaking of the lattice symmetry (*2*, *4*). In the case of the 1T' phase of $MoTe_2$, we think the interlayer stacking fault could give rise to the shear mode. The recent study (*5-7*) indicate the twin structure in the 1T' phase, i.e. the 1T'-I and 1T'-II (*5*) as illustrated in Fig. S3C. The stacking of the 1T'-I and 1T'-II gives rise to an intermediate Td stacking, which could induce the interlayer shear phonon as that in the Td phase. Since the measured unexpected oscillation in the 1T' phase is 0.38 THz, exactly the same as that in the Td phase, we identify the twin structure induced intermediate Td stacking in the 1T' phase by the ultrafast structure response.

A signature of stacking fault and twin structure in the Td phase of $MoTe_2$ could also be identified by ultrafast structure response. As shown in Fig. S4A and B, we observe the disappearance of the intensity oscillation for Bragg reflections belonging to the same family of lattice plane, for example, (350) vs (3-50). Moreover, as the Friedel pairs, (-3-20) shows monotonic decay in contrast to the intensity oscillation of (320). According to the Friedel's law, the intensities of the (hkl) and $\overline{hkl}$ reflections are equal in the electron diffraction pattern. The distinct intensity modulations for Friedel pairs, indicate that the change of the crystal symmetry. The difference map of diffraction intensities in Fig. S4C shows within single diffraction spot, such as (080), (-150) and (-440), the intensity increase in part region and decrease in other part. As the difference map is taken by diffraction patterns at 1.2-1.35 ps (the peak position of the shear mode) subtracting diffraction patterns at 2.6-2.75 ps, the opposite intensity changes within single diffraction spot indicate the opposite phased intensity modulation by the shear mode. In this case, the overall intensity change over the diffraction spot region show quasi-monotonous intensity decay, such as that of (-440) and (-150), in contrast to the strong shear mode modulated intensity changes in (440) and (150). Note that for some spots, the opposite intensity changes region may overlap each other well within single diffraction spot and can not be separated in our experiment. The twin structure (Td-I and Td-II) and the stacking fault in the Td phase of $MoTe_2$ (*5*) are illustrated in Fig. S4D. In the (Td-I)-(Td-II)-(Td-I) stacking, the shear displacement of Td-I, indicated by the yellow arrows, would give rise to the unexpected shear displacement for the intermediate Td-II (the green arrows indicate the expected shear displacement toward a centersymmetric phase). Then the Td-I and Td-II produce oppositely phased intensity oscillations for the same diffraction spot, which coincides with the experimental observation. Therefore, we attribute the anomalous intensity oscillation for diffraction spots from the same lattice family in the Td phase of $XTe_2$ (X=Mo and W) to the twin structure and the interlayer stacking fault.

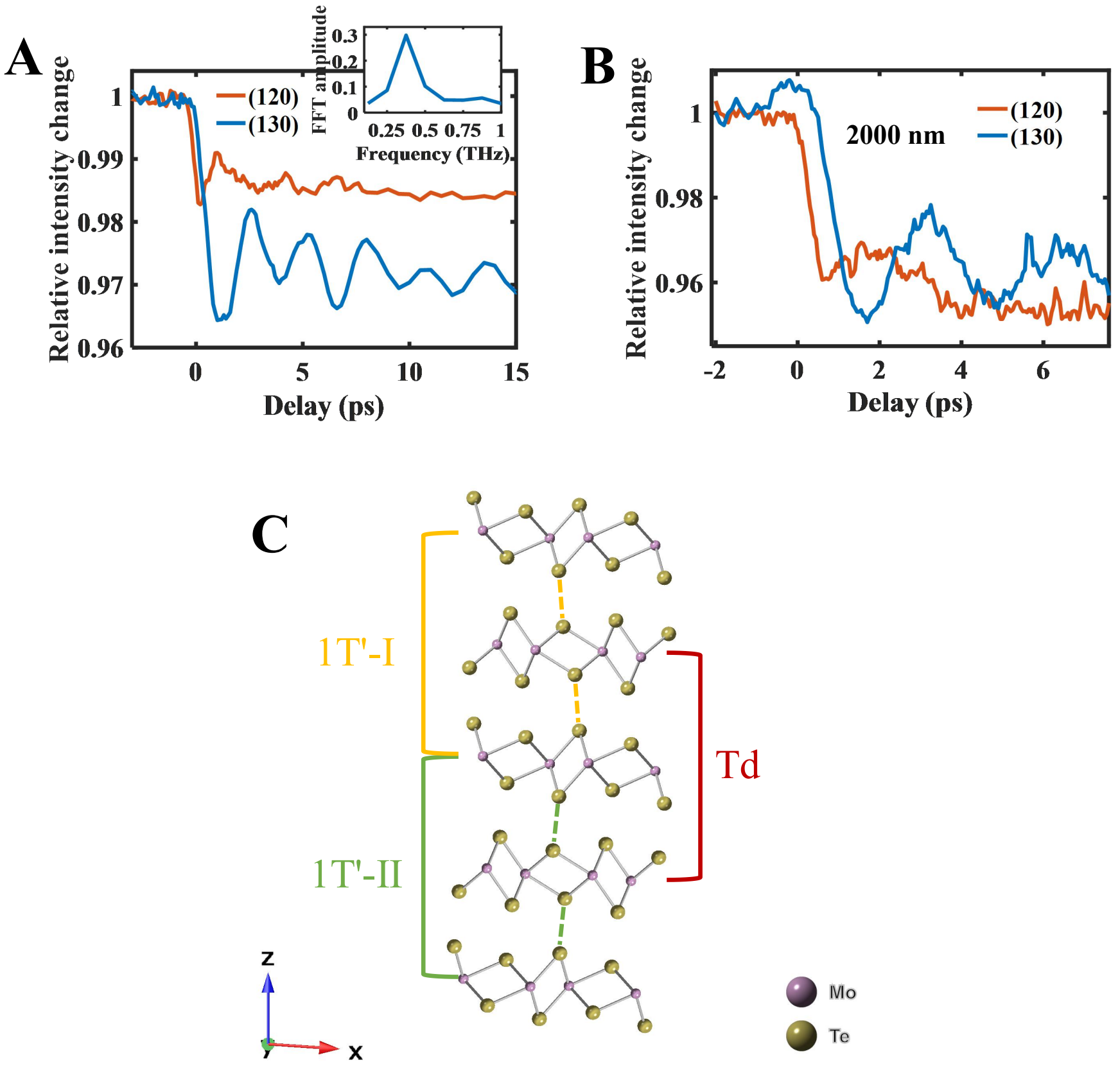

Fig. S3. Intensity oscillation in the 1T' phase of $MoTe_2$ at 295 K. (A) Intensity changes of the (130) and (120) reflection with 550 nm (0.84 mJ/cm$^2$) laser excitation. The inset is the FFT amplitude of the intensity oscillation, indicating a frequency of 0.38 THz. (B) Intensity changes of the (130) and (120) reflection with 2000 nm (13.10 mJ/cm$^2$) laser pump. (C) Illustration of the twin structure in the 1T' phase, i.e. the 1T'-I and 1T'-II. The stacking of the 1T'-I and 1T'-II gives rise to an intermediate Td stacking.

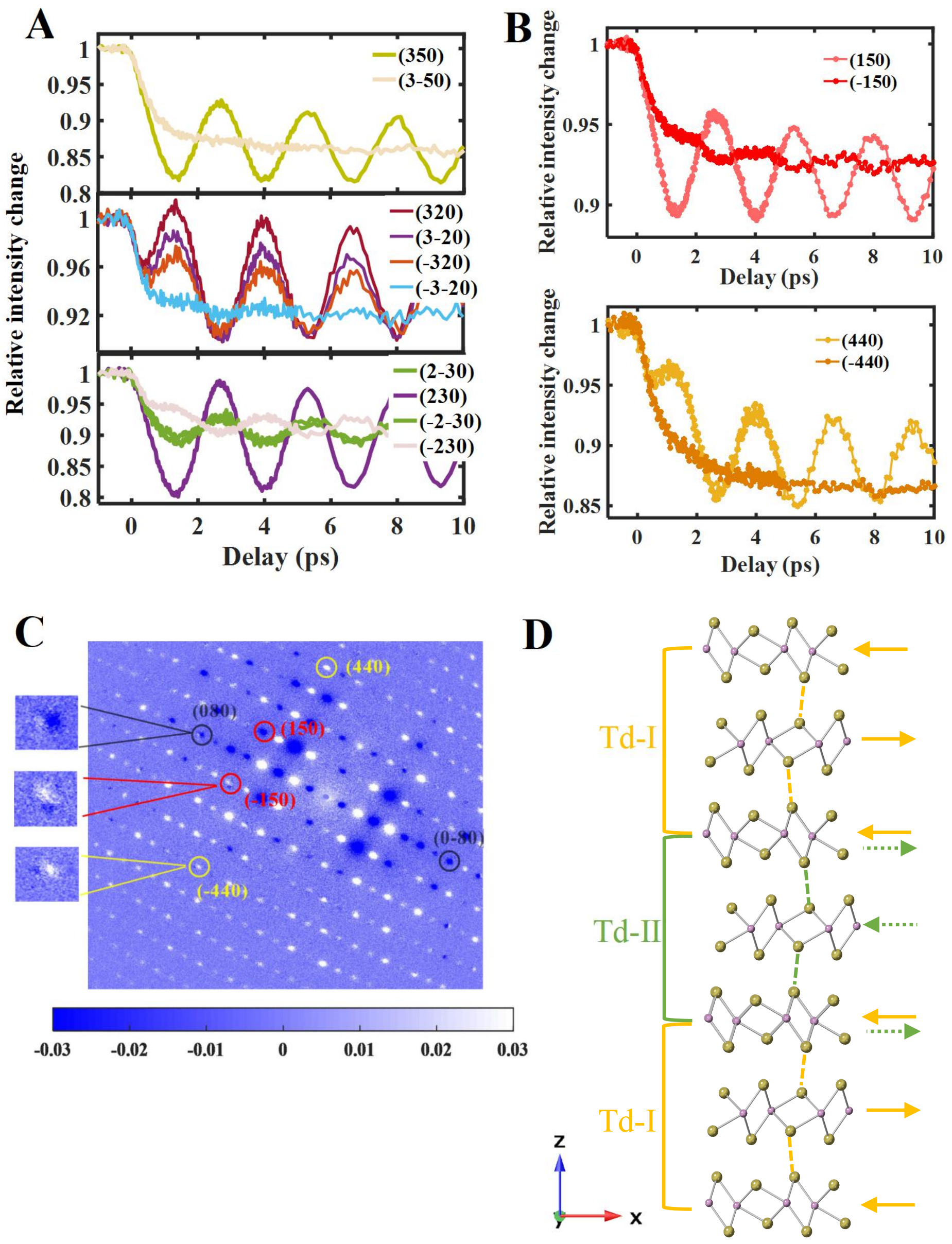

Fig. S4. Stacking fault induced disappearance of the intensity oscillation for Bragg reflections belonging to the same family of lattice plane in the Td phase of $MoTe_2$. The pump laser is 2000 nm and 15.08 mJ/cm$^2$. (C) Difference map of the intensity between the delay points of 1.2-1.35 ps and 2.6-2.75 ps with 2000 nm (13.10 mJ/cm2) laser pump. In single diffraction spot, such as (080), (-150) and (-440), the intensity increase in part region and decrease in other part. (D) Illustration of the twin structure in the Td phase, i.e. Td-I and Td-II, and the stacking of the (Td-I)-(Td-II)-(Td-I). The yellow arrows indicate the displacement direction of the shear mode in Td-I. The dotted green arrows indicate the displacement direction of the intermediate Td-II toward a high symmetry state.

## 3. The fit of the shear mode induced intensity oscillation

We use two methods to fit the intensity oscillation of the Bragg reflections induced by the interlayer shear mode. In the first method, an exponential function and an exponentially decaying cosine function $A*\exp(-t/\tau 1)+B*\cos(\omega t+\varphi)*\exp(-t/\tau 2)$, is used to fit. The results of the fit for several reflections are shown in Fig. 2g in the main text and Fig. S5 (A-E). For the second method, we fit the intensity change at $\geq$ 0.7 ps with an exponentially decaying cosine function $A*\cos(\omega t+\varphi)\exp(-t/\tau)$. In this case, the nonequilibrium dynamics can be excluded and the fitted phase of the oscillation is more precise. The results of the fit is shown in Fig. S5F. With both fit methods, the cosine oscillation of the shear mode is determined.

In the first fit method, the exponential decay part involves three structure responses: the shear displacement beyond the oscillation (i.e. the equilibrium position change), the transient Debye Waller effect and the Mo-Mo displacement. During the fit, the amplitude of the intensity oscillation determines the exponentially decaying cosine function, which in turn influences the exponential decay part. Since the amplitude of the oscillation does not agree with the simulation results very well as shown in Fig. 2f, (the discrepancy may derive from the tilting of the sample and the interlayer stacking fault), so the physical scenario of the exponential decay part in the fit is complex. However, the cosine oscillation of the shear mode is determined and precise.

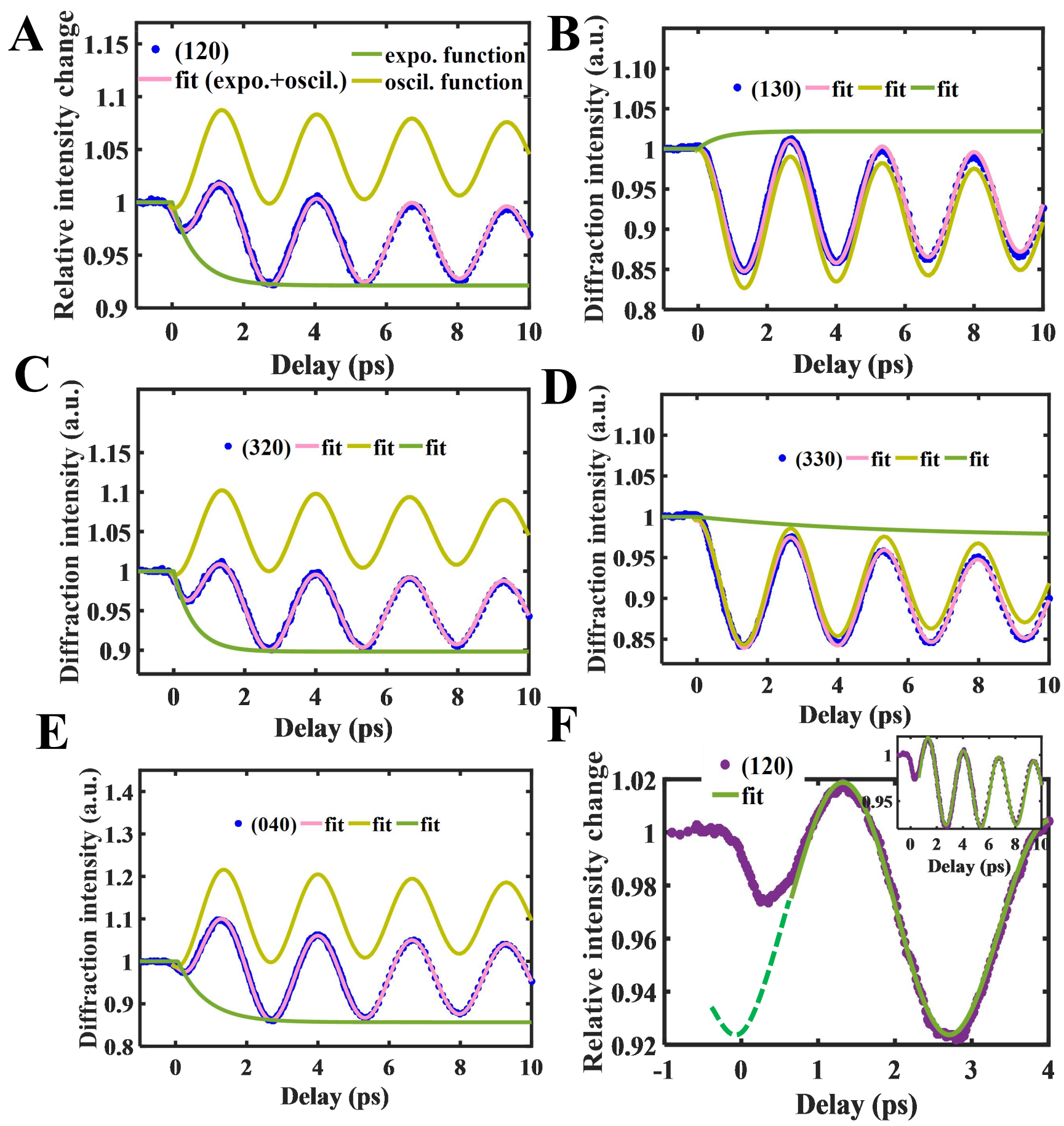

Fig. S5. The fit of the intensity oscillation of Bragg reflections in the Td phase of $MoTe_2$ with 2000 nm laser pump (15.08 mJ/cm$^2$). (A-E) The temporal evolution of the intensity of Bragg reflections and the fit with the same function as that used in the manuscript, i.e. an exponential function and an exponentially decaying cosine function A*exp(-t/τ1)+B*cos(ωt+φ)*exp(-t/τ2). The pink curve is the overall fit. The green curve is the A*exp(-t/τ1) part and the light yellow curve is the B*cos(ωt+φ)*exp(-t/τ2) part, with the parameters determined by the overall fit. (F) The intensity oscillation of the (120) reflection and the fit by an exponentially decaying cosine function A*cos(ωt+φ) exp(-t/τ) (solid green line). The frequency ω and the phase φ of the best fit are 0.38 THz and -0.21 rad. The region of the fit covers 0.7 to 10 ps (see the inset). The dashed green line in the region < 0.7 ps is an extension based on the fit.

## 4. Evaluation of the Debye Waller effect at the equilibrium state

Here, we evaluate the Debye Waller effect at the equilibrium state after femtosecond laser excitation. Since both the shear mode and the intralayer distortion are along the b axis, the intensity change of the (h00) reflections along the a axis indicate the equilibrium of the overall lattice system. The time-resolved intensity change of the {200} and {400} reflection in Fig. S6 suggests the equilibrium of the overall lattice system within ~ 5 ps. The comparable intensity change between 2000 nm (Fig. S6A) and 550 nm (Fig. S6B) laser excitation indicates the same temperature rising after laser excitation. With structure factor calculation, we simulate the intensity change induced by the vibration of Mo and Te atom, as shown in Fig. S6C. The displacement of 0.06 Å induces the intensity change of 5.0% and 17.3% for {200} and {400} reflection, agrees with the measured intensity change of ~5% and ~15%. In Fig. S6D, the intensity change of the (200) and the (400) reflection in the the 1T' phase is shown.

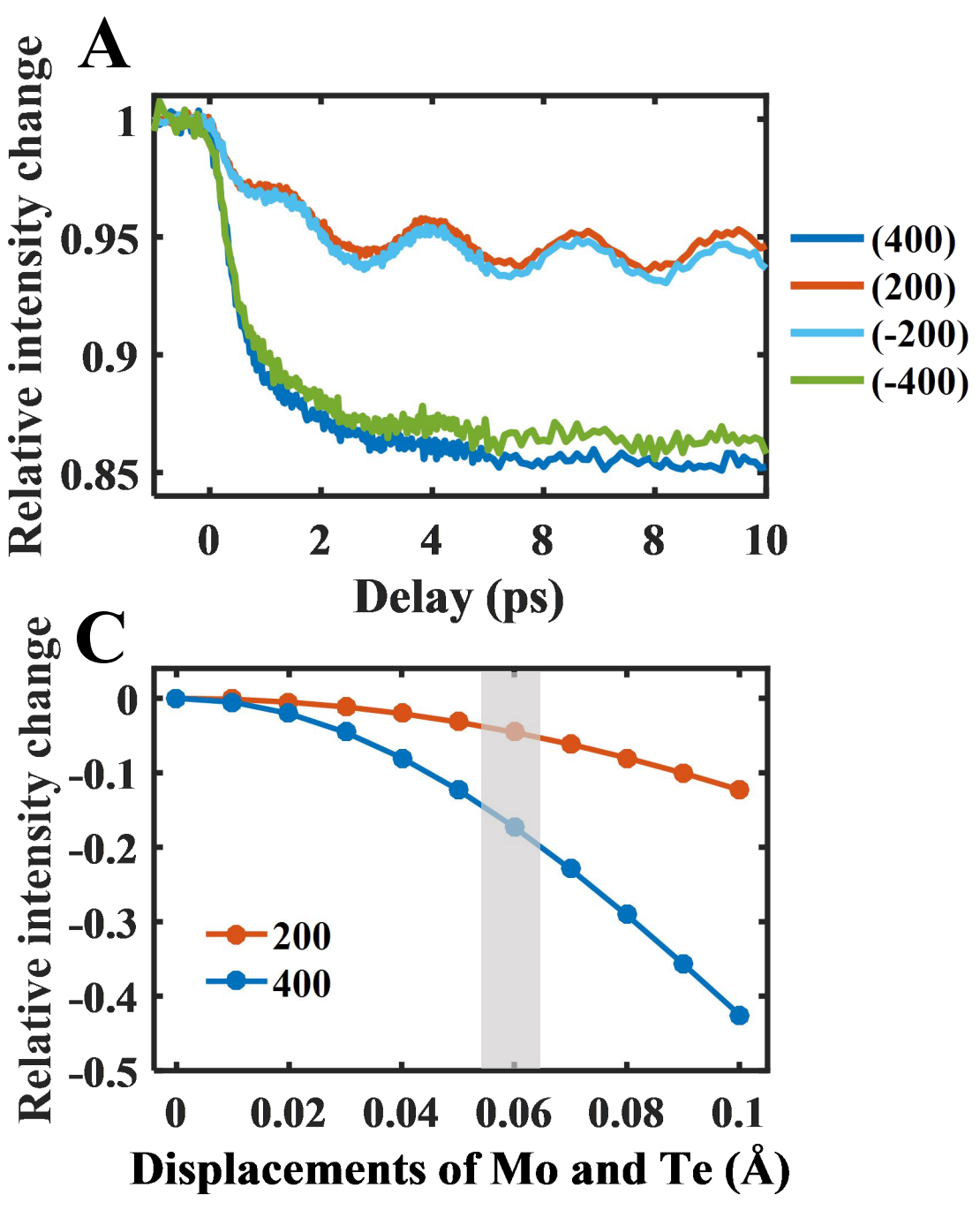

Fig. S6. For the Td phase of $MoTe_2$, the intensity change of {200} and {400} reflection with 2000 nm (15.08 mJ/cm$^2$) (A) and 550 nm (3.81 mJ/cm$^2$) (B) laser excitation. (C) Structure factor calculation of the intensity change by the displacement of Mo and Te atoms in the unit cell. The light rectangular indicates the displacement of ~ 0.06 Å giving rise to the experimental results in (A) and (B). (D) For the 1T' phase, the intensity change of (200) and (400) reflection with 2000 nm (13.10 mJ/cm$^2$).

## 5. Shear displacement induced interlayer structural transition in the Td phase of $MoTe_2$ and $WTe_2$

In Fig. 3 in the main text, we identify, with 550 nm laser excitation (3.81 mJ/cm$^2$), the shear displacement induced interlayer structural transition in the Td phase of $MoTe_2$. With 2000 nm laser excitation (15.08 mJ/cm$^2$), the same intensity change as that in Fig. 3 is shown in Fig. S7A and S7B. The Debye Waller effect for the 550 nm laser excitation (3.81 mJ/cm$^2$) and the 2000 nm laser excitation (15.08 mJ/cm$^2$) is the same (see Fig. S6), so the same interlayer structural transition is produced as expected. For the Td phase of $WTe_2$, the same intensity change of (0k0) reflections as that of $MoTe_2$ is shown in Fig. S7D, suggesting the same interlayer structural transition.

The quantized shear displacement in the interlayer structural transition is based on a systematical analysis. The below is the detailed process. Firstly, we identify the Debye Waller effect induced thermal displacement of 0.06 Å, which is shown in section 4 in Supplementary Materials. Then we combine the thermal displacement and the shear displacement to fit the experimental intensity change of (0k0). With structure factor calculation, the intensity change of (0k0) as a function of the interlayer shear displacement is shown in Fig. S7C. The intensity decrease of (010) is much larger than other (0k0) reflections, which agrees with experimental results. In addition, (010) is the most insensitive reflection to the Debye Waller effect and the intensity change of (010) by the thermal displacement of 0.06 Å is negligible (see Fig. 3B). So we evaluate the shear displacement by the intensity change of (010) and the acquired shear displacement is 0.01 Å. However, for other (0k0) reflections, the calculated intensity change with the shear displacement of 0.01 Å and thermal displacement of 0.06 Å do not show a good agreement with the experimental results (see Fig. 3B), especially for (020) and (040) reflection. Tuning the amplitude of the shear displacement is not working because in experimental result, the intensity of (020) and (040) decreases significantly while in the structure factor calculation, the shear displacement gives rise to intensity enhancement for these two reflections (see S7C). In this case, we consider the contribution from the intralayer Mo-Mo displacement, since in Fig. 4 we have identified the ultrafast intralayer structural transition driven by Mo-Mo displacement. Fig. 4G shows the intensity decrease of (010) by Mo-Mo displacement, therefore, the experimental intensity decrease of (010) should derive from a combination of the shear displacement and the Mo-Mo displacement. We search the minimum standard error (SE) for the intensity change of the (0k0) reflections between the structure factor calculation and the experimental result (k=1 to 5):

$$\mathrm{SE} = \sum_{(0k0)} (\Delta S_{0k0}^2 - \Delta I_{0k0})^2 / \Delta I_{0k0}^2$$

where ΔI is the experimental intensity change and $\Delta S^2$ is the calculated intensity change with a combination of thermal displacement, shear displacement and Mo-Mo displacement. Fig. S7E shows the result of the SE with the search scope of 0.004 Å to 0.008Å for both the shear displacement and the Mo-Mo displacement. The minimum SE labeled by the red circle, i.e. the best fit between the structure factor calculation and experiment results, indicate the shear displacement of 0.005 Å and the Mo-Mo displacement of 0.006 Å.

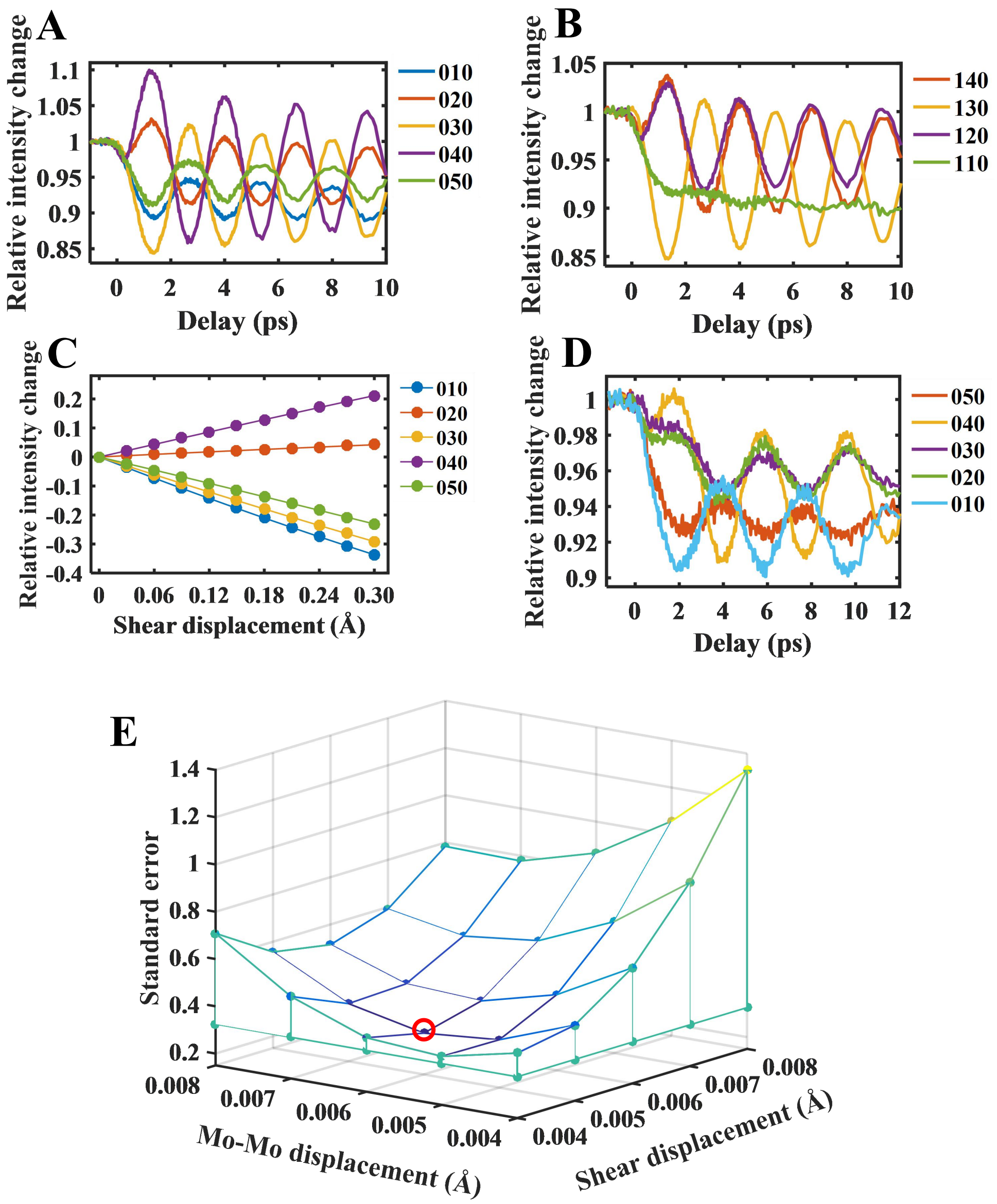

Fig. S7. (A-C) For the Td phase of $MoTe_2$, the intensity change of (0k0) (A) and (1k0) (B) reflections with 2000 nm (15.08 mJ/cm$^2$) laser excitation. (C) Structure factor calculation of the intensity change of (0k0) by the interlayer shear displacement. (D) For the Td phase of $WTe_2$ at 110 K, the intensity change of (0k0) reflections with 2000 nm (13.90 mJ/cm$^2$) laser excitation. (E) The standard error as a function of shear displacement and Mo-Mo displacement. The minimum standard error is labeled by the red circle.

## 6. The same structural response with 550 nm and 2000 nm laser excitation for the Td phase of $MoTe_2$

Here we compare the structural response with 550 nm and 2000 nm laser excitation. In Fig. S8, we observe the 550 nm (3.81 mJ/cm$^2$) and 2000 nm (15.08 mJ/cm$^2$) laser excitation give rise to the same shear mode and the anisotropic intensity change within 0.3 ps. Since Fig. S6 has shown the same Debye Waller effect at equilibrium state for the 550 nm (3.81 mJ/cm$^2$) and 2000 nm (15.08 mJ/cm$^2$) laser excitation, the exciation density is expercted to be the same for these two laser pump condition.Therefore, we conclude that the same structural response excited by these two pump wavelengths.

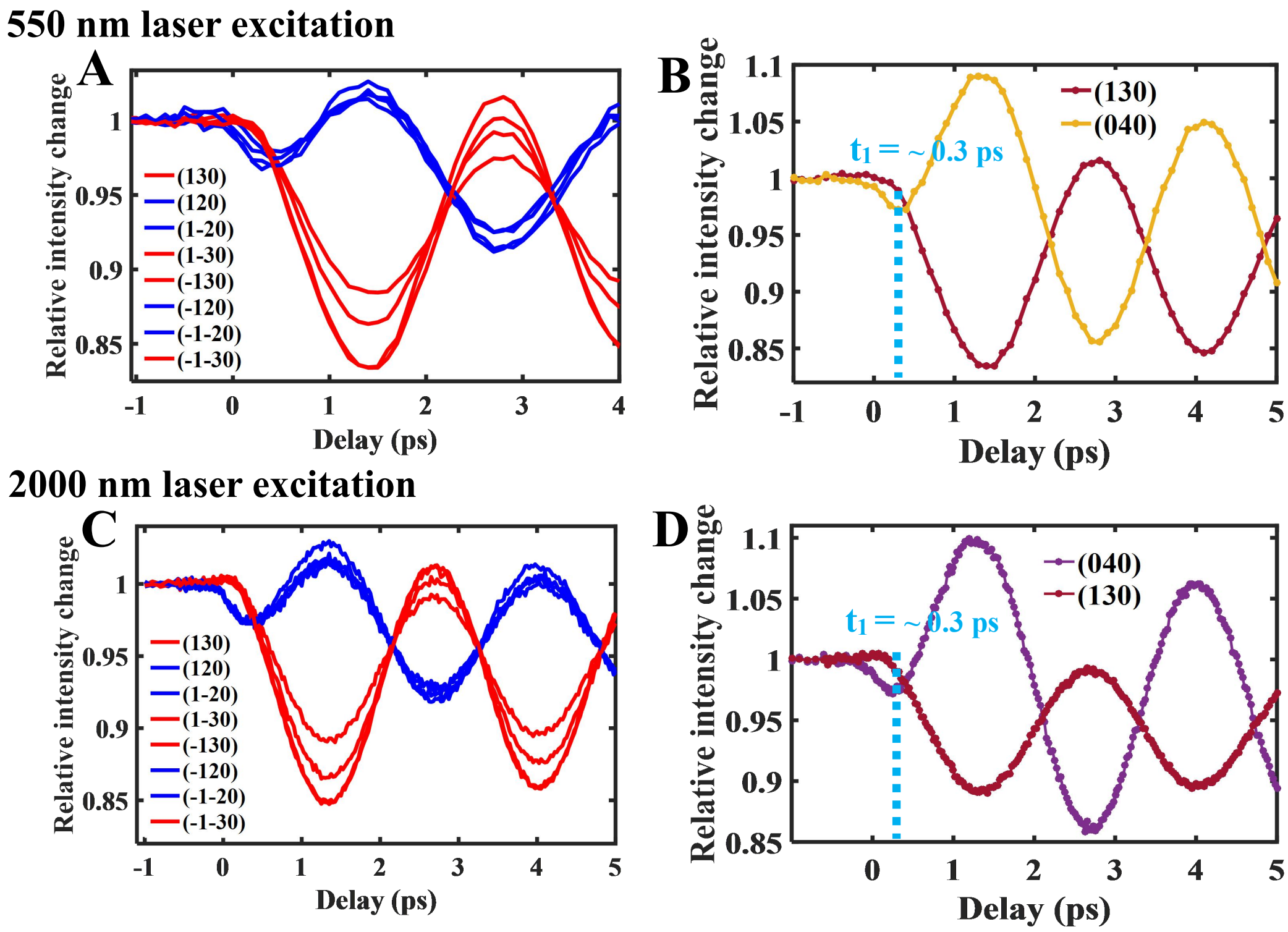

Fig. S8. The structural response with 550 nm (A-B) and 2000 nm (C-D) laser excitation in the Td phase of $MoTe_2$. (A) Time traces for intensity changes of {120} and {130} with 550 nm (3.81 mJ/cm$^2$) laser pump. (B) Time traces for the intensity change of (040) and (130) with 550 nm (3.81 mJ/cm$^2$) laser pump. A prompt decay within 0.3 ps is measured for (040), while a time-delayed intensity decay is observed for (130). (C) Time traces for intensity changes of {120} and {130} with 2000 nm (15.08 mJ/cm$^2$) laser pump. (D) Time traces for the intensity change of (040) and (130) with 2000 nm (15.08 mJ/cm$^2$) laser pump.

## 7. Element-dependent Debye Waller effect to the structural response within 0.3 ps

In the main text, we demonstrate the photoinduced intralayer structural transition,

i.e. the suppression of the intralayer Peierls distortion, within 0.3 ps. Except the model of coherent displacement associated with the intralayer transition, we discuss the Debye Waller of Mo and Te atom to the transient intensity change within 0.3 ps. The relative intensity changes of several Bragg reflections calculated by the displacement of Mo and Te are shown in Fig. S9A and S9B. In experiment results in Fig. 3, the intensity of (040) decreases significantly while the intensity of (330) remains unchanged within 0.3 ps. In contrast to the experiment result, both the Debye Waller of Mo and Te give rise to the larger intensity decay for the (330) than that of the (040). So the Debye Waller effect plays a minor role in the structural response within 0.3 ps. In this work, though we attribute the anisotropic structural response within 0.3 ps to the Mo-Mo bond stretchign and the intralayer structure transition, we can not exclude the minor contribution from the Debye Waller of Mo.

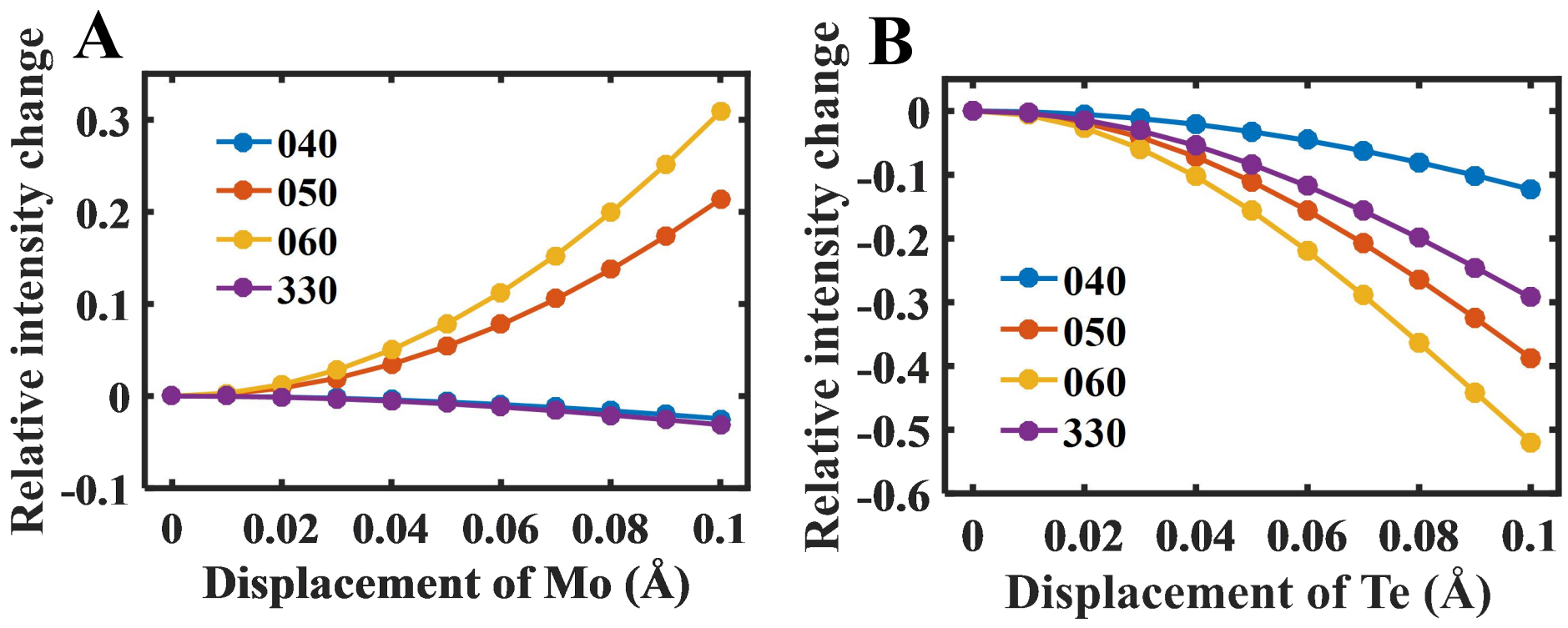

Fig. S9. Structure factor calculation of the relative intensity change induced by atom vibrations in the unit cell. (a) The intensity change of Bragg reflections as a function of displacement of Mo atoms. (b) The intensity change of Bragg reflections as a function of displacement of Te atoms.

## 8. TDDFT-MD simulation of the interlayer and intralayer structure transitions

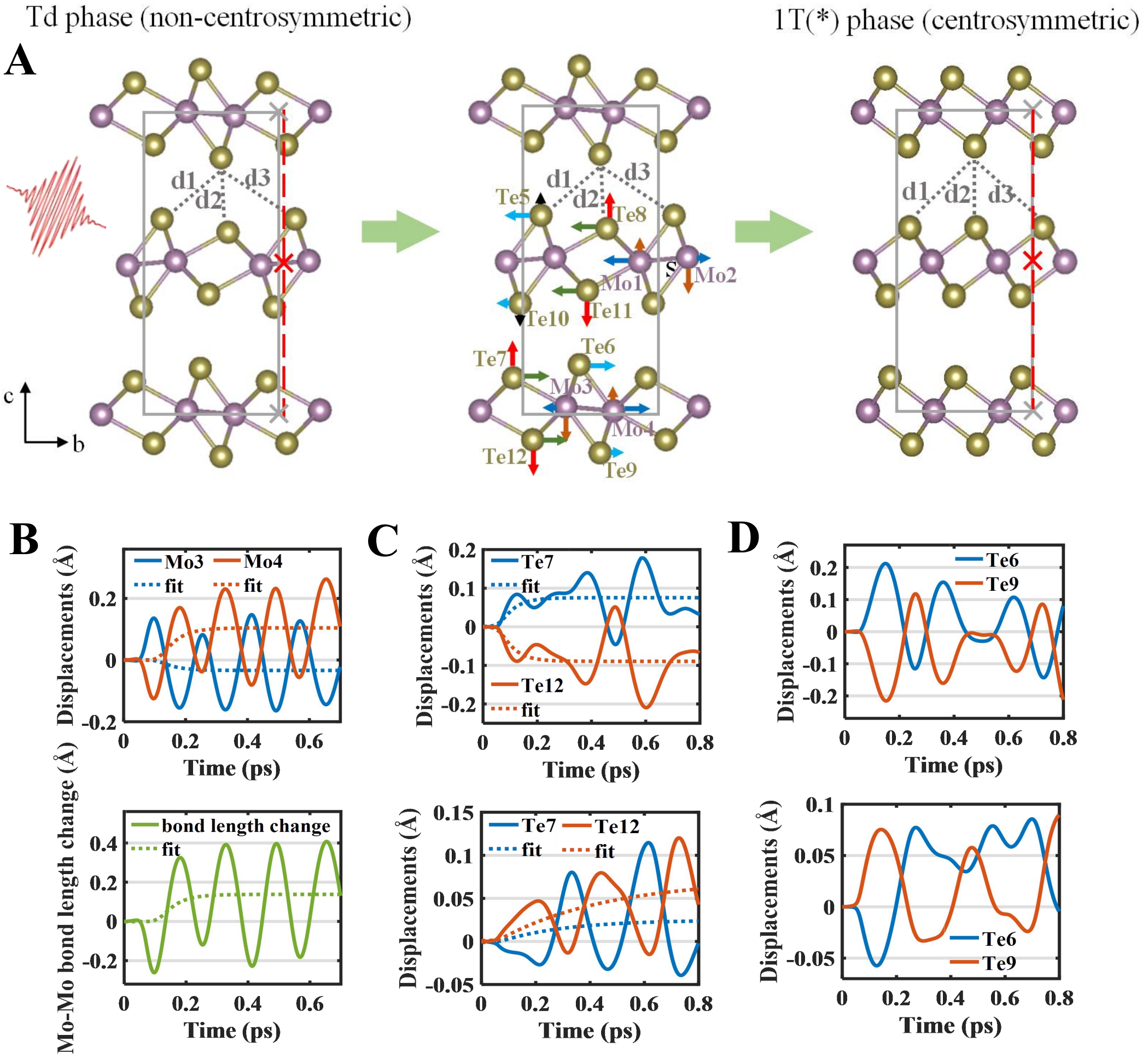

Fig. S10. Simulated atomic motions in the bottom layer of the unit cell. (A) In-plane and out-of-plane movements of Mo and Te atoms indicated by arrows after the laser excitation. (B) (top) The time-dependent displacements of Mo3 and Mo4 along the b axis. (bottom) The time-dependent Mo-Mo bond length change between Mo3 and Mo4. The dotted curves are the monoexponential fit of the simulation results. (C) The time-dependent displacements of Te7 and Te12 along the c axis (top) and the b axis (bottom). The dotted curves are the monoexponential fit of the simulation results.. (D) The time-dependent displacements of Te6 and Te9 along the c axis (top) and the b axis (bottom).

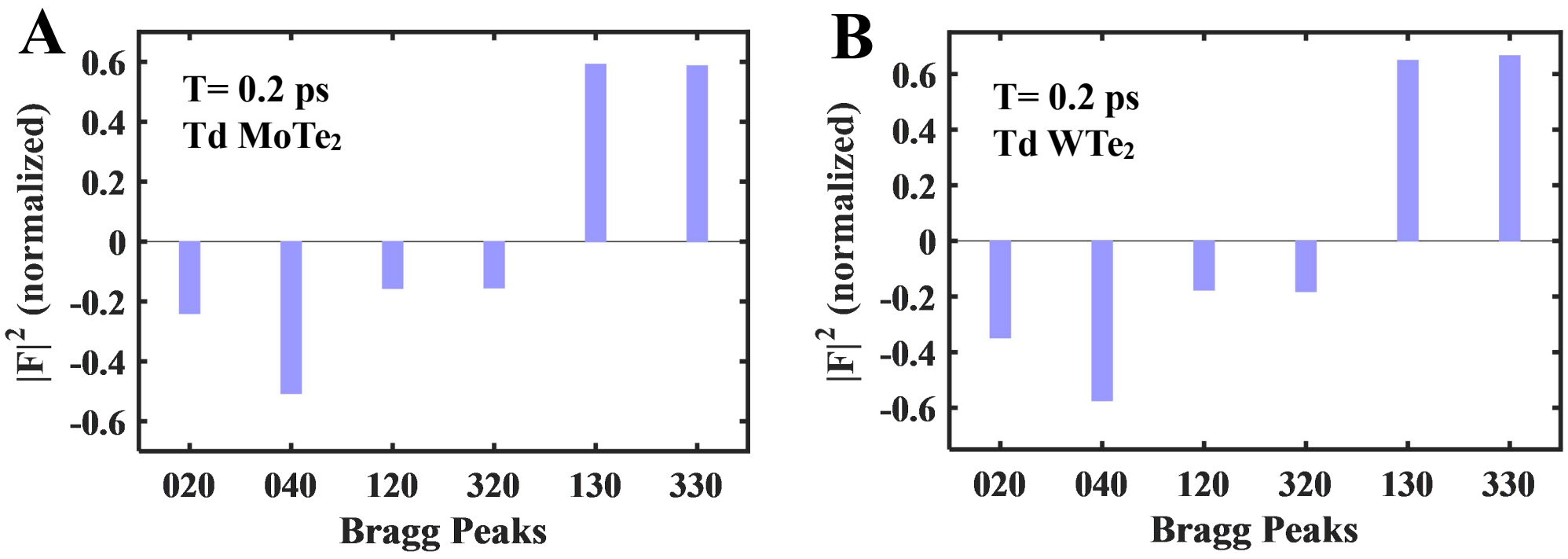

Fig. S11. Bar charts showing calculated intensity changes of Bragg reflections based on simulated atomic displacement at T= 0.2 ps in Fig. S11. The intensity change is calculated by structure factor. For both Td phase of $MoTe_2$ (A) and $WTe_2$ (B), (020), (040), (120), (320) show intensity decrease while (130) and (330) show intensity increase, which agree qualitatively with the experiment results in Fig. 4 and Fig. S2. The discrepancy between the calculated intensity change and the experiment results, such as the percentage and the remained (130) and (330) intensity in experiment VS significant intensity increase of (130) and (330) in calculation, may derive from the inhomogeneous longitudinal excitation due to limited optical penetration depth in experiment.